\documentclass[twocolumn, trackchanges]{aastex63}
\usepackage{natbib}
\usepackage{multirow}
\usepackage{enumerate}
\usepackage{placeins}
\newcommand\tnt{\textsc{\small DYNAMITE}}

\newcommand{\eeri}{\object{e Eridani}}

\received{}
\revised{}
\accepted{}
\submitjournal{AJ}

\shorttitle{e Eridani}
\shortauthors{Basant et al.}
\graphicspath{{./}{}}

\begin{document}

\defcitealias{Feng2017}{F17}
\defcitealias{Pepe2011}{P11}
\defcitealias{2014ApJ...794..133S}{SD14}
\defcitealias{2017AJ....154..160S}{SD17}
\defcitealias{2018MNRAS.478.3025J}{JK18}

\title{An Integrative Analysis of the Rich Planetary System of the Nearby Star e Eridani: Ideal Targets For Exoplanet Imaging and Biosignature Searches}

\correspondingauthor{Ritvik Basant}
\email{ritvikbasant2@email.arizona.edu}

\author[0000-0003-4508-2436]{Ritvik Basant}
\affiliation{Steward Observatory and Department of Astronomy, The University of Arizona, Tucson, AZ 85721, USA}

\author[0000-0001-6320-7410]{Jeremy Dietrich}
\affiliation{Steward Observatory and Department of Astronomy, The University of Arizona, Tucson, AZ 85721, USA}

\author[0000-0003-3714-5855]{D\'aniel Apai}
\affiliation{Steward Observatory and Department of Astronomy, The University of Arizona, Tucson, AZ 85721, USA}
\affiliation{Lunar and Planetary Laboratory and Department of Planetary Sciences, The University of Arizona, Tucson, AZ 85721, USA}



\begin{abstract}
\eeri{}, the fifth-closest Sun-like star, hosts at least three planets and could possibly harbor more. However, the veracity of the planet candidates in the system and its full planetary architecture remain unknown. Here we analyze the planetary architecture of \eeri{} via \tnt{}, a method providing an integrative assessment of the system architecture (and possibly yet-undetected planets) by combining statistical, exoplanet-population level knowledge with incomplete but specific information available on the system. \tnt{} predicts the most likely location of an additional planet in the system based on the Kepler population demographic information from more than 2000 planets. Additionally, we analyze the dynamical stability of \eeri{} system via N-body simulations. Our \tnt{} and dynamical stability analyses provide support for planet candidates g, c, and f, and also predict one additional planet candidate with an orbital period between 549 -- 733 days, in the habitable zone of the system. We find that planet candidate f, if it exists, would also lie in the habitable zone. Our dynamical stability analysis also shows that the \eeri{} planetary eccentricities, as reported, do not allow for a stable system, suggesting that they are lower. We introduce a new statistical approach for estimating the equilibrium and surface temperatures of exoplanets, based on a prior on the planetary albedo distribution. \eeri{} is a rich planetary system with a possibility of containing two potentially habitable planets, and its vicinity to our Solar System makes it an important target for future imaging studies and biosignature searches.
\end{abstract}



\section{Introduction} \label{sec:intro}


Despite the rapid progress in discovering extrasolar planets, the exoplanet population of the solar neighborhood remains mostly unexplored: for example, around the approximately 1,500 stars known within 15~pc \citep[e.g.,][]{Smart2021}, which may host \textit{several thousand} planets, only 159 are known. Given that planets are intrinsically faint, those that are the closest to us (i.e., orbit the closest stars) are set to play key roles in the future of exoplanet studies and, especially, in searches for atmospheric biosignatures \citep[e.g.][]{LUVOIR2019, HabEx2020}. Therefore, it is strategically important to explore nearby planetary systems and to identify those that are the most likely to harbor potentially habitable worlds.
While the information on specific planetary systems almost always remains incomplete and often uncertain, the emerging body of information on exoplanet demographics and on planet formation and evolution can be combined in the future to improve the understanding of individual planetary systems \citep[e.g.,][]{Apai2019}.

The goal of this paper is to study the planetary system of \object{e Eridani} (also known as 82 Eridani, \object{GJ 139}, and \object{HD 20794}; not to be confused with $\epsilon$~Eridani), the fifth-closest Sun-like star, via a novel integrative analysis. Through this analysis, we predict the orbits and properties of yet-undetected planets. Such predictions can help the interpretation of planet candidates identified at low significance levels and can also guide follow-up observations of the planetary system.

At a distance of $6.04$\,pc, \eeri{} is one of the nearest and brightest stars in the sky that is known to host super-Earth planets \citep[e.g.,][]{2018MNRAS.480.2411M}. The first evidence for planets in the \eeri{} system was found in radial velocity data by \citep[][hereafter \citetalias{Pepe2011}]{Pepe2011}, who confirmed two planets and suggested the presence of another. Follow-up observations by \citet[][hereafter \citetalias{Feng2017}]{Feng2017} provided additional evidence for several other planets, with up to six potential planets and candidates known in the system. 


The planetary system around \eeri{} not only has an astronomical significance, but it also has a special place in science fiction and visions of humanity's future.  It is featured in various novels and short stories as a system with habitable worlds which foster intelligent life (c.f., Anderson's \textit{Orbit Unlimited} and Pournelle's short story \textit{He Fell into a Dark Hole}, or Kube-McDowell's story that envisioned it as a home to a primitive human colony). These stories predated the discovery of planets in the \eeri{} system, but modern methods finally enable exploring the planetary architecture of the \eeri{} system and the properties of its worlds.

In this paper, we integrate specific information on the \eeri{} planetary system (constraints on planets and their orbits) with population-level information on exoplanets (exoplanet demographics) to provide a robust assessment of the \eeri{} planetary system. Specifically, we use the \tnt{}\footnote{\url{https://github.com/JeremyDietrich/dynamite}} \citep[][]{2020AJ....160..107D} integrated analysis software package, which was recently successfully utilized in multiple studies (see below). \tnt{} asks the question: If an additional planet exists in a planetary system, what are the probability distributions of its orbital period, eccentricity, inclination, and planet size? 
To answer this question, \tnt{} uses robust trends identified in the Kepler exoplanet demographics data (orbital period distribution, planet-size distribution, etc., based on the $\sim$2,400 exoplanets that form the Kepler population) with specific data for a given single exoplanet system (detected planets and constraints on their orbits and sizes). Based on this information, \tnt{} uses a Monte Carlo approach to map the likelihood of different planetary architectures, also considering orbital dynamical stability and allowing for freedom of statistical model choice. 

By now, \tnt{} has been used successfully in a multitude of studies:
\citet{2020AJ....160..107D} first demonstrated the capabilities of \tnt{} on 45 TESS-discovered transiting planetary systems to predict yet undiscovered planets. In that study, when a known transiting planet -- Kepler-154 f -- was deliberately hidden from the algorithm, \tnt{} not only retrieved the planet but also predicted its size and inclination accurately. When the known non-transiting planet Kepler-20 g was removed from the Kepler-20 system, \tnt{} again recovered the planet. When removing multiple planets from the TOI-174 system, \tnt{} again indicated the likelihood of an additional planet at the location of the purposely removed planets. Even when the periods of other planets are altered within $3\sigma$, the predictions are mostly unaffected by this, and only slightly shifted if all the planetary periods were changed in the same direction. Lastly, when analyzing the HD 219134 system, another planet with a period $\sim$ 12 days is predicted between current planet c and candidate f, along with an additional planet external to planet candidate g (see also Dietrich et al. 2021, submitted).

When \tnt{} was applied to the $\tau$ Ceti system \citep[][]{2021AJ....161...17D}, it provided contextual and statistical evidence for unconfirmed planets b, c, and d, which were found in one earlier study but have marginal evidence over a decade of observations \citep[][]{2013A&A...551A..79T, 2017AJ....154..135F}. In addition, \tnt{} also predicted another yet-unobserved planet in the habitable zone. \cite{2021AJ....161..219H} used \tnt{} on the K2-138 planetary system and accurately predicted planets c and e while analyzing different scenarios in their study. In tests focusing on the inner Solar System, \tnt{} also successfully predicted the positions and sizes of Venus and Earth when one of them was removed from the inner Solar System \citep{Dietrich2022}. While \tnt{} provides a robust way of predicting the likely positions where additional planets could exist in a system, we emphasize that the analysis is based on the available exoplanet demographics data. Therefore, the validity of extrapolating from these results to planet mass $< 1 M_\oplus$ or $> 4 M_\oplus$ and to non-planetary objects (like planetesimal belts) could not be ascertained. Consequently, our analysis and predictions from \tnt{}  are likely to be correct for typical planetary systems but not expected to hold for rare planets or outliers among planetary architectures.

In addition to applying \tnt{} to \eeri{}, in this study we also present an additional module that statistically determines the possible equilibrium and surface temperatures of the planets studied. We also compare the temperature results for the \eeri{} system and the inner Solar System. Though the model we use for this specific task is simple, it still provides some useful insights into the possible temperatures of the planets in the system. 

This paper is organized as follows.  Section~\ref{sec:system} delineates the current knowledge of the \eeri{} system. We describe the \tnt{} package and our implementation of it, the predictive temperature analysis, and the eccentricity analysis in Section~\ref{sec:analysis}.  Section~\ref{sec:results} summarizes the results of our analysis of the planetary  architecture, the equilibrium and surface temperatures of the planets, and the results of our eccentricity analysis.  We discuss the results of the architectural analysis (including the likely number of planets and their eccentricities) in Section~\ref{sec:discussiondynamite}.  In Section~\ref{sec:discussiontemperature}, we motivate our prior on the planetary Bond albedo distribution and then discuss the results of the predicted equilibrium and surface temperature likelihood distributions, as well as the potential for habitable worlds in the \eeri{} system.  Finally, in Section~\ref{sec:bio}, we discuss the potential for habitable worlds in the \eeri{} system.  

\section{The e Eridani planetary System} \label{sec:system}

\subsection{Stellar Parameters}

\eeri{} is a G8V, broadly Sun-like star \citepalias[][]{Pepe2011} located at a distance of $6.0414 \pm 0.0029$ pc \citep {2016A&A...595A...1G, 2018A&A...616A...1G} from Earth, with a luminosity of $0.656 \pm 0.003  L_\odot$ \citep{Sousa2008}. 
Its key stellar parameters are summarized in Table \ref{tab:Stellar}.

\begin{table}[ht]
    {\centering
    \caption{Stellar parameters for \eeri{}}
    \label{tab:Stellar}
    \begin{tabular}{|l|c|r|}
        \hline
        Parameter Name & Value & Ref.\\
        \hline
        Spectral Type & G8V & (a)\\
        Mass [$M_\odot$] & $0.813_{-0.012}^{+0.018}$ & (b)\\
        Radius [$R_\odot$] & $0.92\pm0.02$ & (c)\\
        Luminosity [$L_\odot$] & $0.656\pm0.003$ & (d)\\
        Temperature [K] & $5401\pm17$ & (d)\\
        Distance [pc] & $6.0414 \pm 0.0029$ & (e)\\
        P$_{rot}(R'_{HK}$) \ [d] & $33.19\pm3.61$ & (a)\\
        $[Fe/H]$ [dex] & $-0.40\pm0.01$ & (d)\\
        \hline
        \multirow{3}{*}{Age ($R'_{v_{HK}}$) [Gyr] } & $5.76\pm0.66$ & (a)\\  
        & $> 12.08$ & (c) \\
        & 13.5 & (f) \\
        \hline 
    \end{tabular}
    }
    \\[10pt]
    \textbf{Notes}: (a) \citet{Pepe2011}, (b) \citet{2013ApJ...764...78R}, (c) \citet{2007ApJS..168..297T}, (d) \citet{Sousa2008}, (e) \citet{2018A&A...616A...1G}, (f) \citet{2005ApJS..159..141V} 
\end{table}

\citet{2013ApJ...764...78R} calculated the mass of \eeri{} to be $0.813_{-0.012}^{+0.018} M_\odot$ based on stellar evolution models. The stellar radius of $0.92 \pm 0.02 R_\odot$ was calculated by \citet{2007ApJS..168..297T} through a dense grid of evolutionary tracks. The calculated stellar radius value lies within the range of stellar radii calculated by \citet{2019AJ....158..138S} for this star based on \textit{Gaia} parallaxes and the Stefan--Boltzmann Law. We adopted the derived luminosity, metallicity, and effective temperature values from \citet{Sousa2008}, who carried out a detailed spectroscopic analysis of the HARPS Guaranteed Time Observations (GTO) “high precision” data. We note that a $4.68\%$ higher luminosity than we adopted from \citet[][]{Sousa2008} was calculated by \citet{2018A&A...616A...1G}, but this slightly higher value would not appreciably impact the boundaries of the habitable zone or any of our results. Though \eeri{}'s stellar spectral classification is not yet fully settled, we note that -- based on the values of mass, radius, luminosity, and temperature -- \eeri{} is more likely to be a late G-type star than an early K-type star. However, our analysis relies on the derived luminosity and effective temperature values and not on the spectral type. The age of \eeri{} is likely old, but its exact value remains uncertain:  $5.76$ Gyr is reported by \citetalias{Pepe2011}, an age $> 12.08$ Gyr is found by \citet{2007ApJS..168..297T}, and $13.5$ Gyr is derived by \citet {2005ApJS..159..141V}. The latter two ages have been determined by isochrone fitting, while \citetalias[][]{Pepe2011} used the activity-rotation-age calibration method to calculate the age, following the method given by \citet{2008ApJ...687.1264M}. While such a large disparity in the star's age could affect the potential habitability of the system, the age itself is not utilized in our statistical analysis.
 
\subsection{Planetary system}
\label{S:EEri}
 
In 2011, \citetalias[][]{Pepe2011} announced the discovery of two planets orbiting \eeri{} with orbital periods of 18 days (planet b) and 90 days (planet d). These planets were found using radial velocity (RV) measurements and were identified as likely super-Earths, given their measured $m \sin i$ values. The analysis also found weak evidence for an additional planet, planet candidate c, at a 40-day period.  In a follow-up study by \citetalias[][]{Feng2017}, six different signals were found at periods of 11.9 d (planet g), 18.33 d (confirming planet b), 43.17 d (planet c), 88.9 d (confirming planet d), 147.02 d (planet e), and 331.41 d (planet f). The three additional signals found in the \citet[][]{Feng2017} study resulted from an extended set of observations, more sophisticated data analysis, and superior noise model. Of the six radial velocity peaks reported, three (belonging to planets b, d, and e) were strong while the other three needed further analysis to confirm them.  This newer analysis found marginal evidence for the original planet candidate c from \citetalias{Pepe2011} at a slightly longer 43-day period. 

The planetary signal at an orbital period of 12 days corresponding to planet g is relatively weak, but it consistently appears in the majority of the datasets.  Lastly, the planetary signal at 331.41-day period, planet f, is estimated to lie in the habitable zone of the system, as calculated by \cite{2013ApJ...765..131K, 2014ApJ...787L..29K} for a 5 $M_ \oplus$ planet and 1 $M_ \oplus$ planet around \eeri{}. \citetalias[][]{Feng2017} state that the signal for planet f needs more observation to be confirmed, as the signal is close to the annual sampling period from Earth.  

The proximity of this orbit to a 2:1 resonance with planets b and d, while not actually residing in the resonant orbit, provides some support for its planetary nature, as \citet{2018AJ....156...24M} found that the most common period ratio between neighboring pairs of planets in the Kepler population was $1.9$. Similarly, \citet{2015MNRAS.448.1956S} found that 2:1 resonances between exoplanets are not intrinsically rare. 

The eccentricities of the planets in this system remain poorly constrained. \citetalias{Pepe2011} found that the uncertainties for the eccentricities were of similar order to the values and therefore fixed all eccentricities to 0 (circular orbits). \citetalias{Feng2017} allowed for both circular and Keplerian (non-zero eccentricity) orbits and found that moderate eccentricities ($\sim$0.1-0.3) fit their data best. However, they also state this is likely due to instrumental noise and fitting bias, and so the actual eccentricities are likely lower: Indeed, the orbits of planets d and e would cross and very likely be unstable given the currently reported values. Later in this paper we explore the possible range of eccentricities that lead to stable systems. 

We summarize the relevant parameters of the six planets and planet candidates in Table~\ref{tab:planets}.
 
\begin{table*}[ht]
    \centering
    \caption{Planet and Planet Candidate Parameters}
    \label{tab:planets}
    \begin{tabular}{|l|c|c|c|c|l|}
        \hline
        \textbf{Name} & \textbf{Period} & \textbf{Semi-major axis } & \textbf{Eccentricity} &\textbf{msin i} & \textbf{Notes} \\
        \textbf{ } & \textbf{[days]} & \textbf{[au]} & & \textbf{[$M_\oplus$]} & \\
        \hline
        g & $11.86_{-0.02}^{+0.01}$ & $0.095_{-0.001}^{+0.001}$ & $0.20_{-0.19}^{+0.15}$ & $1.03_{-0.30}^{+0.49}$ & Candidate: Reported in F17\\
        b & $18.33_{-0.02}^{+0.01}$ & $0.127_{-0.001}^{+0.001}$ & $0.27_{-0.22}^{+0.04}$ & $2.82_{-0.80}^{+0.10}$ & {Confirmed: Reported in P11, confirmed in F17}\\
        c & $43.17_{-0.10}^{+0.12}$ & $0.225_{-0.003}^{+0.002}$ & $0.17_{-0.16}^{+0.10}$ & $2.52_{-0.83}^{+0.52}$ & Candidate: Reported in both P11 and F17\\
        d & $88.90_{-0.41}^{+0.37}$ & $0.364_{-0.004}^{+0.004}$ & $0.25_{-0.21}^{+0.16}$ & $3.52_{-1.01}^{+0.58}$ & {Confirmed: Reported in P11, confirmed in F17}\\
        e & $147.02_{-0.91}^{+1.43}$ & $0.509_{-0.006}^{+0.006}$ & $0.29_{-0.18}^{+0.14}$ & $4.77_{-0.86}^{+0.96}$ & {Confirmed: Reliable detection in F17}\\
        f & $331.41_{-3.01}^{+5.08}$ & $0.875_{-0.010}^{+0.011}$ & $0.05_{-0.05}^{+0.06}$ & $10.26_{-1.47}^{+1.89}$ & Candidate: Reported in F17 \\
        \hline
    \end{tabular}
    \\[10pt]
    \textbf{Note}: These planetary parameters have been collected from \cite{Feng2017}.  
\end{table*}

Although not used in our study, for completeness we note that the \eeri{} system was also found to host a debris disk, revealed by infrared excess detected with the \textit{Spitzer Space Telescope} \citep{2012MNRAS.424.1206W}.  Follow-up observations with the \textit{Herschel Space Observatory} marginally resolved the disk emission and the disk's inner and outer radii were estimated to be around 20 and 30 au, with the disk relatively depleted beyond 30 au \citep{2015MNRAS.449.3121K}.  Modeling of the disk spectral energy distribution argued for a cold disk (T$_{eq}=80^{+70}_{-30}~K$) with blackbody disk radius about 10 au, based on disk temperature \citep{2015MNRAS.449.3121K}.  The debris disk provides evidence for a planetesimal belt exterior to the known planets, similarly to the Solar System's Kuiper Belt. In the future, studies of this planetesimal belt may provide constraints on the outer planetary system in the future.

\section{Analysis} \label{sec:analysis}

\subsection{The \tnt{} Package} \label{sec:dynamite}

Here we briefly explain the functionality of the \tnt{} software package and the mode that we use in our study. As stated earlier, \tnt{} amalgamates various statistical distributions identified in the Kepler exoplanet population data and combines that general information with the specific data of a given planetary system. It then integrates this with a dynamical stability criterion and tests the probability density functions (via Monte Carlo injections) for orbital period, planetary radius, and planetary inclination. Finally, \tnt{} also considers the orbital stability of the different planetary architectures. This combined information allows the algorithm to estimate the likelihoods for the parameters of one additional unknown planet in the system. The basic assumptions we used to constrain the probability distribution in \tnt{} are:

\begin{enumerate}
    \item The period, planet radius, and orbital inclination are independent of each other. 
    
    \item The orbital periods considered range from 0.5 days to 2 years, corresponding to the approximate range for which strong constraints exists on planetary architectures and planet properties from the Kepler mission-discovered exoplanet population. 
    
    \item The planet radius range considered extends from 0.5 to 5 $R_\oplus$. Here the lower limit comes from the Kepler population and the upper limit empirically contains a large majority of the TESS multi-planet systems. \citep[][]{2020AJ....160..107D}.  Planets larger than 5 $R_\oplus$ tend to be rare and may have distinct orbital period, size, and mass distributions \citep[e.g.,][]{2019MNRAS.490.4575H}.
    
    \item The range of allowed planetary inclinations is 0$^{\circ}$-- 180$^{\circ}$.
    
    \item The dynamical stability is determined either via N-body simulations and the spectral fraction analysis from \citet[][]{2020AJ....160...98V} or via the mutual Hill parameter for neighboring pairs of planets $\Delta_c \geq 8$ \citep[][]{Chambers1996}, depending on user input.
\end{enumerate}

For a detailed discussion of these assumptions, please see \citet[][]{2020AJ....160..107D}. In our analysis, we used the period ratio model from \citet[][]{2018AJ....156...24M}, which assumes a broken power law for location of the first planet in period space (with the break at 12 days) and equal period ratios between pairs of planets. Therefore, this model inserts planets roughly symmetrically in gaps as a consequence of this chosen distribution, with the highest probability occurring when the period ratios are closest to the Kepler population period ratio mean of 1.9.

The inclination distribution, in the case of no known planetary inclinations, is isotropic but limited to non-transiting values for each planet (as the planets are known not to transit). The planetary radius distribution is a Lognormal distribution around a central cluster, which is fit to the radii of the known planets, or if the planet radii are unknown, from the assumed radii derived from the known mass or m sin i values and the \citet[][]{Otegi2020} M-R relationship). After injecting a planet, we calculate the dynamical stability parameter and compare it to the threshold value to see if the system would stably exist with that planet addition.

Through 10$^6$ Monte Carlo iterations, we build up the likelihoods for the additional planet parameters. While the results are reliant on the specific data for the given system being accurate, they are not sensitive to the exact values. We utilize the default parameter set for each planet from the NASA Exoplanet Archive\footnote{\url{https://exoplanetarchive.ipac.caltech.edu/}}, and the uncertainties on most of these parameters are low enough to enable a robust analysis. We explore the impact of uncertainties on the \tnt{} output predictions in Section \ref{subsec:RV} and Appendix \ref{sec: additionalhypotheses}. 

\subsection{Planetary Configuration Hypotheses} \label{sec:dynamethods}

Given the uncertain nature of three planet candidates in the system, we proceeded in our analysis by constructing hypotheses that explore all possible combinations of the planet candidates being genuine or not. For each hypothesis, we provide \tnt{} predictions. As explained later, we found that the predicted planetary orbits that results from these different hypotheses are very similar. In the main body of the current paper we present four hypotheses, while in the Appendix  \ref{sec: additionalhypotheses} we show that results of the additional hypotheses. 
The hypotheses are constructed by starting from the three confirmed planets, and by adding planet candidates one by one, in the order of decreasing evidence for their existence. Thus, our first hypothesis (H[a]) includes only the three confirmed planets (b, d, e), and our second hypothesis (H[b]) adds to this set the planet candidate for which evidence is the most convincing (planet c, detected in two papers). Then, in our third hypothesis (H[c]), we included planet candidate f as a genuine planet (stronger candidate signal detected by \citetalias{Feng2017}). Finally, we assume that all planet candidates are genuine planets (Hypothesis H[d]).

We note that, in the above sequence of hypotheses, predictions from each of the hypotheses motivates the next one, as described (i.e., H[a] predicts a planetary architecture consistent with H[b], etc.). As discussed in \S\ref{sec:discussiondynamite}, our predictions for the planetary architecture are robust and do not display strong dependence on the presence or absence of any single planet. Our analysis of the planetary architecture assesses different scenarios to reflect the yet unsettled nature of the candidate planets.  We organize these possibilities into four hypotheses, corresponding to four different potential states of the planetary system given current constraints: 

\begin{itemize}

 \item \textbf{Hypothesis H[a]:} Only the three planets b, d, and e are taken as known planets.  None of the three planet candidates (g, c, f) are genuine detections.
    
 \item \textbf{Hypothesis H[b]:} In addition to the three planets (b, d, and e), planet candidate c is assumed to be a genuine planet.  Planet candidates f and g are not considered to be real detections.
    
 \item \textbf{Hypothesis H[c]:} Planets b, c, d, e, and f are all genuine detections, but the peak corresponding to planet g is not a detection.

 \item \textbf{Hypothesis H[d]:} All six planet detections (g, b, c, d, e and f) proposed by \citetalias{Feng2017} are real planets.

\end{itemize}


We do not fix the inclinations of the planets and planet candidates, as those values are unknown due to the inability to measure them with RV observations alone, although we can constrain the inclination upper bound via the determination that these planets do not transit. We give the entire range of inclination possibilities to \tnt{}, thus lowering the chances of biasing the model towards any specific planetary architecture type. The orbital eccentricity of the planets was also input based on multiplicity-dependent statistical distributions from the Kepler population \citep{2020AJ....160..276H}. Non-zero eccentricities and the statistical distribution of orbital eccentricities are explored further in \citep[][]{Dietrich2022}.

We also predict the planet radius and mass of additional planets for each hypothesis.  We use a clustered planet radius model from \citet[][]{2019MNRAS.490.4575H} that fits the currently known planet radii to a Lognormal distribution, from which the values are drawn for the Monte Carlo iterations of the additional planets. These planet radii are also converted into mass values via the mass-radius relationship from \citet[][]{Otegi2020}, specifically using the ``volatile-rich" power-law for planets in the mass range $5-25 M_\oplus$, with planets below $5 M_\oplus$ assumed to be rocky and planets above $25 M_\oplus$ assumed to contain a large gaseous envelope.  We expect the planets in this system of sufficient mass orbit far enough away from \eeri{} that they would accrete and retain a significant gaseous envelope, instead of it being stripped away via photoevaporation \citep[e.g.,][]{Owen2017}.

\subsection{Surface Temperature Assessment} \label{sec:tempmethods}

With 3-6 planets possible in the planetary system of \eeri{}, the potential equilibrium and surface temperatures of these worlds are of great interest.  As available data are scarce, we will follow a statistical approach to explore the potential range of temperatures on these worlds.  We do so by expanding our \tnt{} modeling framework with statistical predictions of the equilibrium and approximate surface temperatures. Our analysis should be taken only as an initial exploratory assessment, until more constraining data on the planets are available; nevertheless, as shown later in the manuscript, our analysis provides interesting insights into the planets' possible natures. In the following text we will describe the methodology used to provide the initial, exploratory assessment of the potential range of planetary temperatures for the predicted planets.

The calculation of $T_{eq}$ is based on two assumptions: first, the planet (with its atmosphere) is a blackbody, and second, the complete system is in radiative equilibrium with its surroundings. This is a very good approximation of the average temperature (i.e., disregarding day/night side variations) for an airless body (e.g., the Moon, Mars), and is accurate within 10\% for planets without an atmospheric greenhouse effect. For \eeri{}, we adopt the ${a_{planet}}$ (the semi-major axis) and $P_{planet}$ (orbital period) values for the planets and planet candidates from the recent RV study by \citetalias[][]{Feng2017}.

For every orbital period from the Monte Carlo iterations from \tnt{}, we draw an albedo from a normal distribution with mean $\mu = 0.3$ and standard deviation $\sigma = 0.1$ and calculate the radiative equilibrium temperature of the planets via Eq.~\ref{eq:t} ($A_{B}$ is the Bond albedo, $T_{star}$ and $R_{Star}$ are the surface temperature and radius of the star respectively, and $a_{planet}$ is planet's semi-major axis).
\begin{equation} \label{eq:t}
T_{eq} = T_{star}(1-A_{B})^\frac{1}{4}\sqrt{\frac{R_{star}}{2a_{planet}}},
\end{equation}
A detailed discussion of our choice of a Gaussian distribution for the Bond albedo with the above mentioned parameters is given in Appendix~\ref{sec:albedoselection}.

We note here that the available data on exoplanet albedos is very sparse, and there remain uncertainties on the mass-radius conversion. These may lead to systematic offsets in our albedo model. Nevertheless, our results are not very sensitive to the exact albedo values, in part because our assumed albedo distribution is relatively wide. For example, the assumption that the albedo of planet b in the \eeri{} system is 0.6, instead of 0.3, results in a temperature decrease of $\sim13\%$ from the equilibrium temperature calculated with albedo of 0.3. Moreover, as the albedo distribution we adopted is relatively broad and is derived using the bodies in the Solar system, few objects are expected to lie outside of this range.

Perhaps the most important limitation of our simple model is the lack of an atmosphere and the associated greenhouse effect.  To address this, we calculated the surface equilibrium temperatures using a leaky greenhouse atmosphere model \citep{2008MP}, wherein we assume a single-layered atmosphere (which approximates to a blackbody) that absorbs a fraction {$\alpha$} of the radiation from the planet. Our single-layer leaky greenhouse model is a reasonable first approximation for planets without heavy atmospheres and very strong greenhouse warming.  Additionally, more comprehensive models could provide more realistic predictions of the surface temperatures, but would -- in turn -- also require assumptions on the atmospheric compositions and cloud properties which are unconstrained for most exoplanets.  Therefore, a simple model with just one free parameter is a better choice. We use a normal distribution for $\alpha$ ($\mu = 0.5$ and $\sigma = 0.1$) so that nearly $100\%$ of the values lie between 0 and 1. Due to the isotropic nature of the thermal emission, the atmosphere then radiates equal amounts of energy towards the planet as well as away from it. The energy balance of the stellar irradiation, surface emission, and single-layer greenhouse absorber can then be described as follows:

\begin{equation}\label{eq: 2}
T_{surface} = \left(\frac{2}{2 - \alpha}\right)^\frac{1}{4}T_{eq},
\end{equation}

where $T_{\text{surface}}$ is the surface temperature.  

To report the most likely equilibrium and surface temperatures for planet candidates, we select the median along with the 16th and 84th percentiles of orbital period distribution from \tnt{} and convert those into the mean temperatures and their standard deviations. Thus, our temperature measurements mostly depend on the uncertainties in the predictions of the location of planet candidates by \tnt{}; the uncertainties of the orbital period are (relatively) much smaller than the uncertainties typical to the other orbital parameters or planetary masses. The results are plotted as a 2-D histogram in Figure~\ref{fig:dynamitetempcompare} and Figure~\ref{fig:dynamitetemp}.

For statistically predicting the equilibrium and possible surface temperatures of planets (b, d, and e) in the \eeri{} system, we plot another histogram using a linear orbital period distribution centered at the planet's proposed orbital period and extends as per the uncertainties proposed by \citetalias[][]{Feng2017}.  With the orbital period prior, we use the same approach of calculating relative likelihood of surface temperatures with normally drawn albedos between 0 and 1 with $\mu = 0.3$ and $\sigma = 0.1$.

\subsection{The Eccentricity Assessment}\label{subsec:eccentricityanalysis}

With multiple planets on close orbits, the analysis of the long term stability of the \eeri{} system can provide important insights. \citetalias[][]{Feng2017} acknowledged that the planetary eccentricities in their best-fit multi-planet radial velocity models were, in fact, moderate-to-high, which may cause instability in the system. Our goal here is to evaluate the stability of the system with the eccentricities provided by \citetalias[][]{Feng2017} and provide  updated bounds on the eccentricities of the planets in the \eeri{} system. In order to analyse the stability of the system, we use REBOUND \citep[]{rebound, reboundmercurius} software package to run N-body simulations with two different eccentricity distributions: (1) a Normal distribution for each planet's eccentricity with parameters provided by \citetalias[][]{Feng2017} and (2) a Lognormal distribution for each planet's eccentricity with parameters provided by \citet{2020AJ....160..276H}.

The dynamical stability criterion that we use for each randomly drawn eccentricity checks the system for four different conditions:

\begin{enumerate}
    \item The orbits of neighboring planets crossing each other before the N-body integration begins.
    
    \item Close encounters and ejections during N-body integration.
    
    \item The orbits of neighboring planets crossing each other during N-body integration.
    
    \item Spectral Fraction \citep{2020AJ....160...98V}, as explained below.
\end{enumerate}

The spectral fraction analysis executes short N-body integrations ($\sim 5\times 10^{6}$ orbits of the innermost planet) to predict the long-term ($\sim 5\times 10^{9}$ orbits of the innermost planet) stability of the system. This is done by determining the ``spectral fraction'', or the fraction of frequencies having power $\geq5\%$ of the peak frequency in the power spectrum of the angular momentum deficit (AMD, the difference between the total angular momentum and planar circular angular momentum). If the spectral fraction is $\leq5\%$, then the system is assumed to be likely stable long-term. Else, the system is assumed to be unstable long-term.

After assessing the likely planetary architectures for \eeri{} using \tnt{}, we identify two limiting architectures for the system -- Hypotheses H[a] and  H[d] -- which drive the loosest and the tightest constraints on the planetary eccentricities. Therefore, we explore these two specific architectures -- the 3-planet system and the 6-planet system -- to help us analyze the potential planetary eccentricities for this system. For each simulation, we iterate over 1000 randomly drawn eccentricities from the chosen distributions mentioned above and report the number of stable and unstable iterations. For every stable and unstable iteration, we calculate the combined average eccentricity of all the planets involved in the run. We then report the combined mean eccentricity for the planets in the system and associated standard deviation of all the cases that resulted in a stable and an unstable configuration in the simulation.

\section{Results} \label{sec:results}

\subsection{System Architecture Analysis}

\begin{figure*}[ht]
    \centering
    \includegraphics[width=0.9\linewidth]{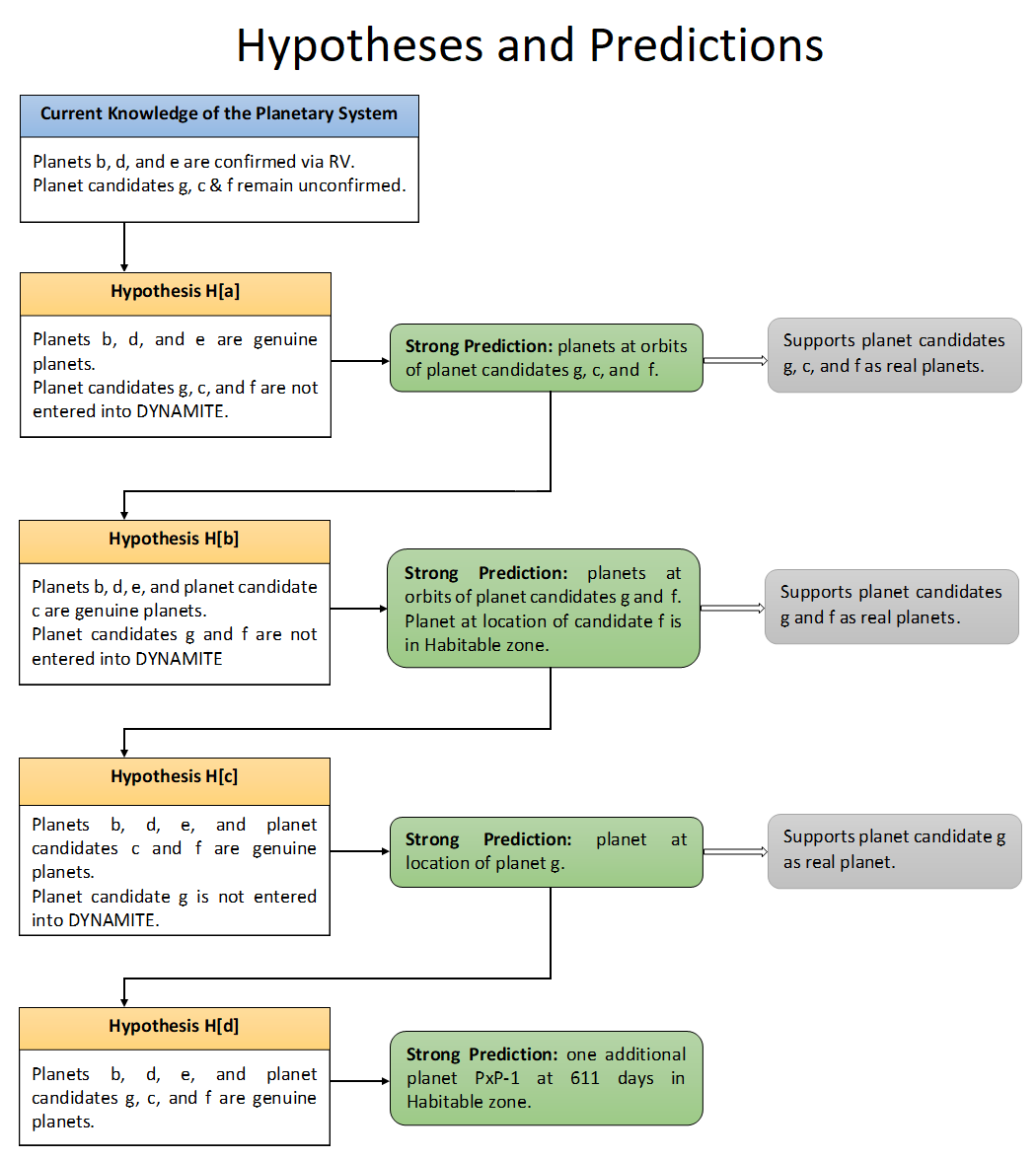}
    \caption{Hypotheses and Analysis Predictions for the planetary System \eeri{}.}
    \label{fig:flow_hypothesis}
\end{figure*}

\begin{figure*}[ht]
    \centering
    \includegraphics[width=1.92\columnwidth]{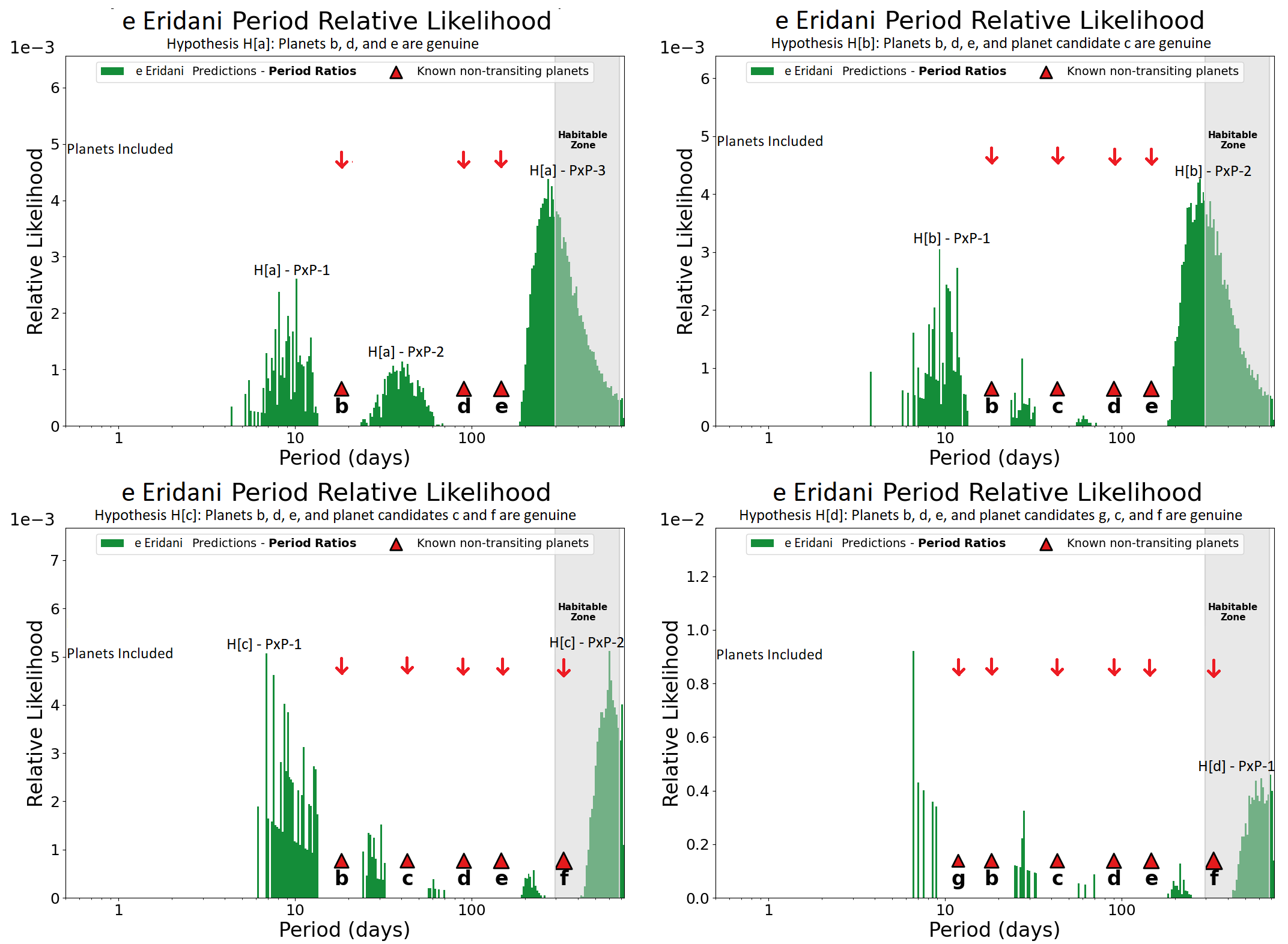}
    \caption{The upper left plot shows the \tnt{} analysis without planet candidates g, c, and f.  The upper right plot shows \tnt{} analysis without planet candidates g and f.  The lower left plot shows \tnt{} analysis without planet candidate g while the lower right plot shows the \tnt{} analysis with all the planets and planet candidates.}
    \label{fig:HD20794_1}
\end{figure*}

\paragraph{\textbf{Hypothesis a: planets b, d, and e}}

In our first hypothesis, Hypothesis H[a], we only placed the planets b, d, and e in the system. We found two roughly equally significant peaks in the probability distribution, H[a]--PxP--1 (i.e., Predicted eXoPlanet 1 in Hypothesis H[a]) at 10.9 {[7.7, 12]} ($16^{th}$ and $84^{th}$ percentiles) days (interior to planet b) and H[a]--PxP--2 at 40.7 [34.2, 52.4] days between planets b and d. A third, slightly larger probability distribution peak is identified as H[a]--PxP--3 at 271 [269, 520] days, exterior to planet e.

\paragraph{\textbf{Hypothesis b: Accepting planet candidate c}}

We considered planet candidate c as a real planet in the system.  In our analysis of this hypothesis, we recovered peaks H[a]--PxP--1 and H[a]--PxP--3 as H[b]--PxP--1 and H[b]--PxP--2, respectively.  In addition to these two peaks, we also found two small regions of non-zero probability in the period space, between planets b and c, and planets c and d, but the injection probability of any additional planet in those gaps is relatively low.  Therefore, this hypothesis predicts two planets, H[b]--PxP--1 and H[b]--PxP--2 at 10.9 and 271 days, respectively.  

\paragraph{\textbf{Hypothesis c: Accepting planet candidates c and f}}

In Hypothesis H[c], we considered planet candidate f and c to be genuine planets along with planets (b, d and e), due to the support given by H[b]--PxP--2. In Hypothesis H[b], the distribution of potential orbital periods of the outermost planet is wider as compared to other predictions because \tnt{} only has one neighbouring planet to use in constraining the orbit. However, as this predicted orbit also matches the period of planet candidate f closely, this prediction provides support for existence of planet candidate f under the assumption of lower planetary eccentricities. In this case, we found a new  peak for H[c]--PxP--2, the most exterior likelihood peak, at 611 [549, 733] days, and we also recovered H[b]--PxP--1 as H[c]--PxP--1.  The newly found peak, H[c]--PxP--2, hints at the presence of an additional planet in the system in the habitable zone of this system, near the outer edge.  However, as this peak is again an extrapolation, lying beyond the longest-period confirmed planet, the predictions are here less constraining.  Furthermore, we also found 3 regions between planets b and c, c and d, and e and f where the probability for an exoplanet to exist is comparatively low but non-zero.  

\paragraph{\textbf{Hypothesis d: Additional planets}}

Finally in Hypothesis H[d], we assumed (given the support from the previous hypotheses) that all the planet candidates (g, c and f) are real and added them in the system along with the planets (b, d and e). In Hypothesis H[c], the inner peak is better constrained due to two reasons: (1) the presence of an inner known planet and (2) the broken power law on the period followed by the first planet from the orbital period model. As the peak of the broken power law occurs at $\sim$12 days, very near the orbital period of planet candidate g, this provides support for the genuine nature of planet candidate g. In our analysis of this planetary architecture, we retrieve the H[c]--PxP--2 peak at 611 days as H[d]--PxP--1, as well as four small regions in the period space with low injection probability interior to planet g and between planets b and c, c and d, and e and f. While there is no direct evidence for the existence of the predicted exoplanet H[d]--PxP--1 yet, our prediction is based on robust statistics from thousands of Kepler multi-planet systems. By construction, DYNAMITE predicts the most likely position in orbital period space for  an additional planet to exist. When the system has low probability of finding an additional planet interior to the outermost system (i.e., when the system is dynamically packed like the \eeri{} system with all candidates treated as genuine planets), then the most likely orbit for an additional planet will be beyond the orbit of the outermost known planet. We also note that this prediction is only valid under the assumption that the eccentricities of the planets in the \eeri{} system are lower (see section~\ref{sub:eccana}) than previously reported.

For such an additional planet in the \eeri{} system, we find the most probable planet radius to be $1.4^{+0.525}_{-0.365}R\oplus$, from fitting a Lognormal distribution to the most likely planet radii given the known planet minimum masses and the ``volatile-rich" mass-radius relationship from \citet[][]{Otegi2020}. From this, we find the most likely planet mass to be $2.91^{+5.79}_{-1.88}M\oplus$. These values are estimated by \tnt{}, given the minimum masses of the planets, the above mass-radius relationship, dynamical stability considerations with the six-planet system architecture, a uniform prior on the inclination assuming the planet is likely not transiting, and the distribution in planet radii assuming size clustering in a system \citep[e.g.,][and see \S\ref{sec:dynamethods}]{2019MNRAS.490.4575H}.

\subsection{Potential Equilibrium and Surface Temperatures} \label{subsec: temp}

We estimated the equilibrium temperature of planet b to be $641 \pm 23$~K, planet d to be $379 \pm 14$~K, and planet e to be $320 \pm 12$~K. The uncertainty in the equilibrium temperature of planets is the standard deviation of the temperature distribution calculated via a linear prior on the orbital period, with limits on the orbital period uncertainty provided by \citetalias[][]{Feng2017}. With the help of the orbital period probability distribution from \tnt{}, we estimated the equilibrium temperature of planet candidate g to be $791 \pm 42$~K, planet candidate c to be $486 \pm 26$~K, and planet candidate f to be $241 \pm 17$~K. We calculated these temperatures assuming that only the planets b, d, and e are genuine, but even if a peak in the distribution is not occupied by a genuine planet, the other peaks remain unaffected.

\begin{figure*}[ht]
    \centering
    \includegraphics[width=1.92\columnwidth]{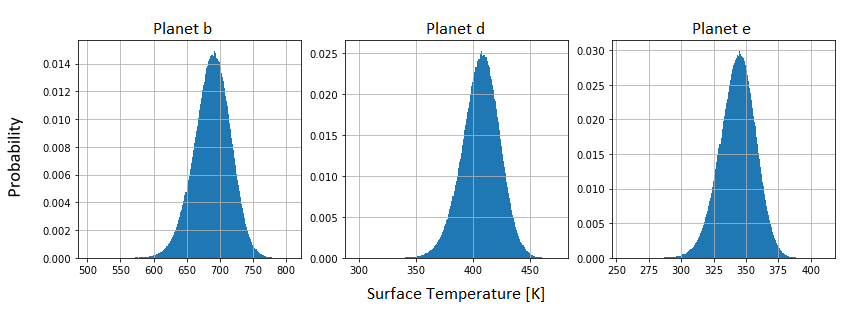}
    \caption{Likelihood of surface temperatures for planets b, d, and e in the system \eeri{}.}
    \label{fig:planetwisebde}
\end{figure*}

Radiative equilibrium temperatures, however, may underestimate surface temperatures of bodies with significant atmospheres due to additional heat trapping by greenhouse gases. To account for this, we expanded our findings of equilibrium temperatures and calculated the possible surface temperatures of the planets and planet candidates in the system (see \S\ref{sec:tempmethods}). These surface temperatures are calculated assuming thin atmospheres and low greenhouse warming, in this case a simple one-layer ``leaky greenhouse'' model. Based on this model, we estimate the possible surface equilibrium temperature of planet b to be $689 \pm 27$~K, planet d to be $407 \pm 16$~K, and planet e to be $344 \pm 14$~K.  These results are shown in Figure~\ref{fig:planetwisebde}.  For the planet candidates, we calculate the surface temperature of planet candidate g to be approximately $849 \pm 52$~K, planet candidate c to be $519 \pm 29$~K, and planet candidate f to be $259 \pm 19$~K.  These results can be found in Figure~\ref{fig:dynamitetempcompare} and Figure~\ref{fig:dynamitetemp}.  The uncertainties in the surface temperature are calculated in the same way as calculated for equilibrium temperature.

In Hypothesis H[d], where we assumed that all six planetary signals found by \citetalias[][]{Feng2017} are genuine planets, we found a significant peak we call H[d]--PxP--1 at an orbital period of 611 days.  When performing our temperature analysis on this planet candidate, we found that the equilibrium temperature of this planet would likely be $196 \pm 9$~K.  We also explored the potential greenhouse warming on this planet and estimate its surface temperature to be $210 \pm 10$~K.  

\subsection{Eccentricity Analysis}\label{sub:eccana}

In the following we report the fraction of stable and unstable orbital configuration iterations for four different hypotheses -- system architecture hypotheses H[a] and H[d], each with both Normal and Lognormal eccentricity distributions. In essence, the normal distribution represents the solution reported by \citet{Feng2017}, while the Lognormal eccentricity distribution represents an agnostic approach based on typical planetary eccentricities. For each of our four cases, we report the combined mean eccentricity and its standard deviation. The results are also shown in Figure~\ref{fig:ecc_ana} in Appendix~\ref{appendix:eccana2}.

\paragraph{Hypothesis H[a] - Normal Distributions}
Assuming that the \eeri{} system has only 3 planets -- b, d, and e -- and a Normal eccentricity distribution centered on the values reported by \citetalias{Feng2017} resulted in only $1.2\%$ of iterations being stable configurations via the criteria established in \S\ref{subsec:eccentricityanalysis}. Planets in the stable configurations have a mean eccentricity of $0.16 \pm  0.02$, whereas the planets in the unstable configurations have a mean eccentricity of $0.24 \pm 0.04$. 

\paragraph{Hypothesis H[a] - Lognormal Distributions}
Under the above planetary architecture assumption but with a Lognormal eccentricity distribution from SysSim \citep{2020AJ....160..276H} instead, we observe that $98.3\%$ of our N-body integrations yield stable configurations. Using this distribution, planets in the stable configurations have a mean eccentricity of $0.05 \pm  0.02$ while planets in the unstable configurations have a mean eccentricity of $0.14 \pm 0.03$.

\paragraph{Hypothesis H[d] - Normal Distributions}
When we assumed that the \eeri{} system consists of 6 planets -- g, b, c, d, e, and f -- and that these planets have a Normal eccentricity distribution, we find that only one out of thousand iterations resulted in stable configurations. The eccentricities for each planet in the stable iteration were 0.069, 0.127, 0.161, 0.137, 0.15, and 0.064 for planets g, b, c, d, e, and f respectively. All the other 999 iterations resulted in unstable configurations.

\paragraph{Hypothesis H[d] - Lognormal Distributions}
Lastly, for a 6 planet system with the planets having eccentricities as per Lognormal distribution resulted in $98.9 \%$ of iterations being stable configurations. The mean eccentricity for all the planets in stable configurations was $0.026 \pm 0.008$, while it was $0.065 \pm 0.018$ for all the planets in unstable configurations.

\begin{figure*}[ht] 
    \centering
    \includegraphics[width=1.92\columnwidth]{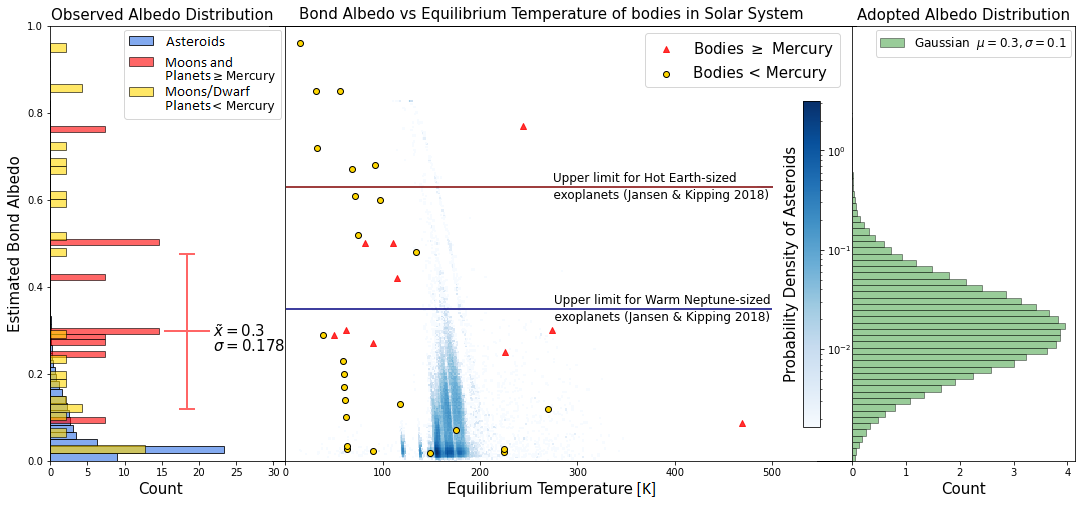}
    \caption{Relationship between Bond albedos and calculated radiative equilibrium temperatures for planets, satellites, and asteroids in the Solar System.  For the adopted Gaussian albedo prior ($\mu=0.3$, $\sigma=0.1$), roughly $69\%$ and $99.8\%$ of the drawn albedo value fall below $A_{B} = 0.35$ and $A_{B} = 0.63$ respectively.  Thus our adopted albedo prior is consistent with known Solar System and exoplanet constrains.}
    \label{fig:Heatmap}
\end{figure*}

\section{Discussion - System Architecture} \label{sec:discussiondynamite}


\subsection{Sensitivity to Radial Velocity Solutions \label{subsec:RV}}

In terms of planetary system-specific information, our analysis uses the results of planet detections from \citet[][]{Pepe2011} and \citet[][]{Feng2017}. These detections are results of multi-parameter fits to the observed radial velocity modulations. To some extent, the results of such orbital fits depend on the underlying assumptions. A particular challenge in fits to radial velocity curves in multi-planet systems is the inter-dependence of the planets' orbital parameters (for example, fitting for three planets may yield planets with different eccentricities than a fit assuming four planets). While these effects have been considered carefully in \citet[][]{Feng2017} -- by  increasing step-by-step the complexity of fits and relying on Bayesian Information Criteria to identify the preferred solutions -- it is not trivial how such correlated uncertainties propagate into our analysis.
Therefore, we will now discuss to what extent  uncertainties in these studies and choices made by their authors may impact our results. The specific questions we will explore are as follows: \textit{How does the number of planets detected impact our results?} and \textit{How do the radial velocity models (and their setup) impact our predictions?}

Ultimately, our analysis is based on the veracity of the three confirmed planets (b, d, and e, see Table~\ref{tab:planets}). In essence, these confirmed planets anchor the orbital architecture of the system. These three planets have been identified independently in two studies (b, and d) or detected in one state-of-the-art study at high significance level (e). Therefore, given the current knowledge on the system, the existence of these planets is safe to assume. As uncertainties remain on the orbital parameters of these planets, it is possible that their orbital periods, minimum mass, or eccentricities is somewhat different than currently inferred from the data. However, our results show that the planetary system architectures derived are insensitive to small- to moderate changes in the orbital parameters of these planets. 

We note, however, that should one of these three planets turn out to be highly eccentric (not consistent with current orbital fits, \citealt[][]{Feng2017}) or a spurious detection, then our analysis must be repeated and results may change. While such a development is unlikely in the face of the available data, it is prudent to discuss how our results may be affected by the existence and specific orbital parameters of the yet unconfirmed planet candidates (g, c, and f). Here, our analysis can provide a clear demonstration that the parameters or existence of these planet candidates does not have an effect on our prediction. This is the case because through the comprehensive set of hypotheses we tested (see \S\ref{sec:dynamethods} and Appendix~\ref{sec: additionalhypotheses}), every combination of the genuine or spurious nature of these planet candidates has been explored. These repeated, independent assessments all converged to the same general planetary architecture, regardless of the existence or properties of the three candidate planets. Therefore, we conclude that our analysis is insensitive to the presence of the planet candidates, but would be affected if one of the three confirmed planets would, in fact, turn out to be a spurious detection.

\subsection{Eccentricity Bounds}

Our eccentricity analysis relies on the Normal and Lognormal eccentricity distributions parameters estimated by \citetalias[][]{Feng2017} and \cite{2020AJ....160..276H}, respectively. We tested the \eeri{} planetary system for the 3-planet and 6-planet architectures for both the above mentioned eccentricity distributions. In our analysis, we first integrated every planet's orbit for $\sim 5\times10^{6}$ orbits of the innermost planet and then predicted its long-term stability ($\sim 5\times10^{9}$ innermost planet's orbits) using spectral analysis. For the specific case of the \eeri{} system, our long term analysis thus corresponds to roughly 1.2 Gyr in the future. Any iteration that resulted in a collision, ejection, crossing of orbits during integration, or a high spectral fraction was considered as an unstable configuration.  Stable pairs of planets with crossing orbits can exist, such as Neptune--Pluto (due to their 3:2 resonance) or potential planet pairs with relatively high mutual inclination and equal mass ratios \citep[][]{Rice2018}, but in this system we are very unlikely to encounter such a situation (as these planets are likely non-resonant). 

Our analysis suggests that even if the system has only 3 planets (b, d, and e), with the eccentricity parameters calculated by \citetalias[][]{Feng2017} it is still highly unlikely that the system will remain dynamically stable over 1.2 billion years. If the system instead has 6 planets, then the likelihood of the system being dynamically stable over 1.2 billion years reduces even further to nearly zero. For this analysis, we used the minimum masses and the semi-major axes of the planets and planet candidates as calculated by \citetalias[][]{Feng2017}. When we used the Lognormal parameters from \citet{2020AJ....160..276H} for our stability analysis, we found that more than 98\% of iterations were stable for both system architectures tested. Thus, these results convey that the chances of the system being dynamically stable for at least a billion years with any planetary architecture are highly likely.

\citetalias[][]{Feng2017} acknowledged that their eccentricity solutions were high and that factors like instrumental noise could be one of the reasons for such high eccentricities. Given that the \eeri{} system is thought to be older than 5.76 billion years, arguably the system is now -- and has been -- in a dynamically stable configuration. Our stability analysis, however, shows that stability is only possible if the planets indeed have lower eccentricities than those reported by \citetalias[][]{Feng2017}. We estimate that if the \eeri{} system  has three planets -- b, d, and e -- and the planet period and minimum mass parameters are close to those calculated by \citetalias[][]{Feng2017} then each planet's eccentricity would be of order $0.05 \pm 0.02$. Similarly, if \eeri{} is a 6-planet system -- with planets g, b, c, d, e, and f -- then the most likely eccentricity for each planet would be of order $0.026 \pm 0.008$.

We note that if the eccentricity vales are updated and better constrained in the future, then our analyses would have to be repeated. We also want to point out that our results from \tnt{} analysis are valid under the assumption that the planets, and the candidates if they exits, will have lower eccentricity values than those estimated by \citepalias[][]{Feng2017} in order for the planetary system to be dynamically stable.

\section{Exploration of Planetary Temperatures} \label{sec:discussiontemperature}

{Here we explore the potential temperatures for the predicted exoplanets, considering the ranges of their predicted semi-major axes as well as a simple statistical prior on planetary Bond albedos.  As little is known yet about the planets and predicted planets discussed here, the following discussion aims to provide only a simple, initial exploration of the potential temperatures of the planets in order to further our understanding of the potential natures of these worlds. The results presented in \S\ref{subsec: temp} are an extension of the Stefan-Boltzmann Law. We motivate and discuss our choice of Bond albedo prior in Appendix~\ref{sec:albedoselection}. Later, we also discuss the possibility of habitable planets in the \eeri{} system.}


\subsection{Equilibrium and Surface temperatures}

For planets b and d, we estimated the equilibrium temperatures to be in the ranges of $618-664$~K and $365-393$~K, respectively.  These ranges are $\pm 1\sigma$ from the median values.  Our results are fairly consistent with those of \citetalias[][]{Pepe2011}, stated above.  For planet e, we estimate its equilibrium temperature to be in the range $308-332$~K.  We estimate the equilibrium temperature of planet candidate g to be in the range $735-819$~K, planet candidate c to be $455-509$~K, and planet candidate f to be $224-258$~K.  Our results for equilibrium temperature of planet candidate c is consistent with that of \citetalias[][]{Pepe2011}.

\begin{figure*}[ht]
    \centering
    \includegraphics[width=1.92\columnwidth]{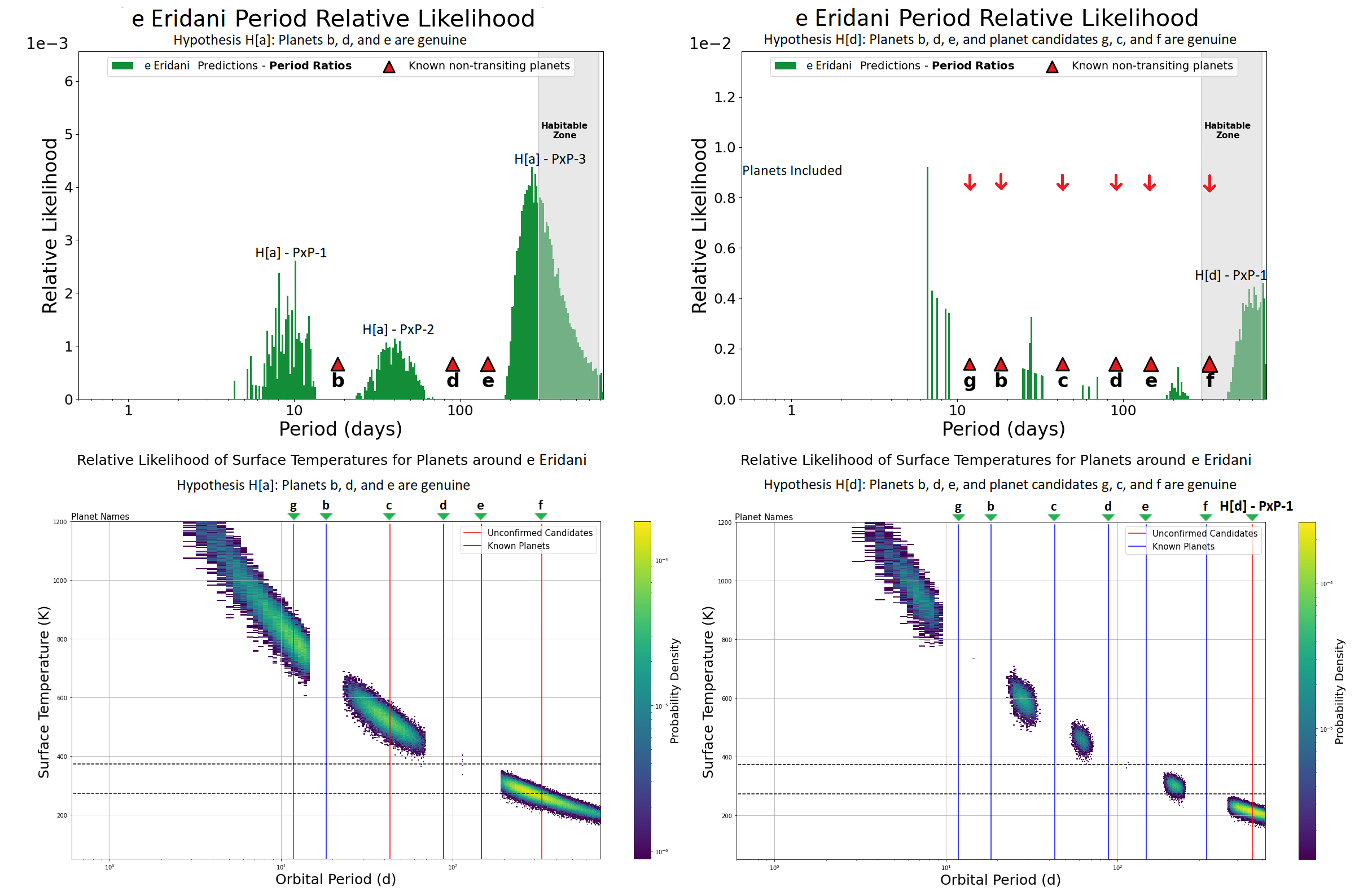}
    \caption{This plot shows a side-by-side comparison of the \tnt{} planet prediction and the new module of \tnt{} that allows statistical prediction of surface temperatures of planets and candidates for Hypotheses H[a] and H[d]. Eccentricities for all planets were set to zero for calculating these temperatures, but utilizing the expected low eccentricity values typical of these multi-planet systems does not significantly affect the temperature.}
    \label{fig:dynamitetempcompare}
\end{figure*}

\begin{figure*}[ht]
    \centering
    \includegraphics[width=1.88\columnwidth]{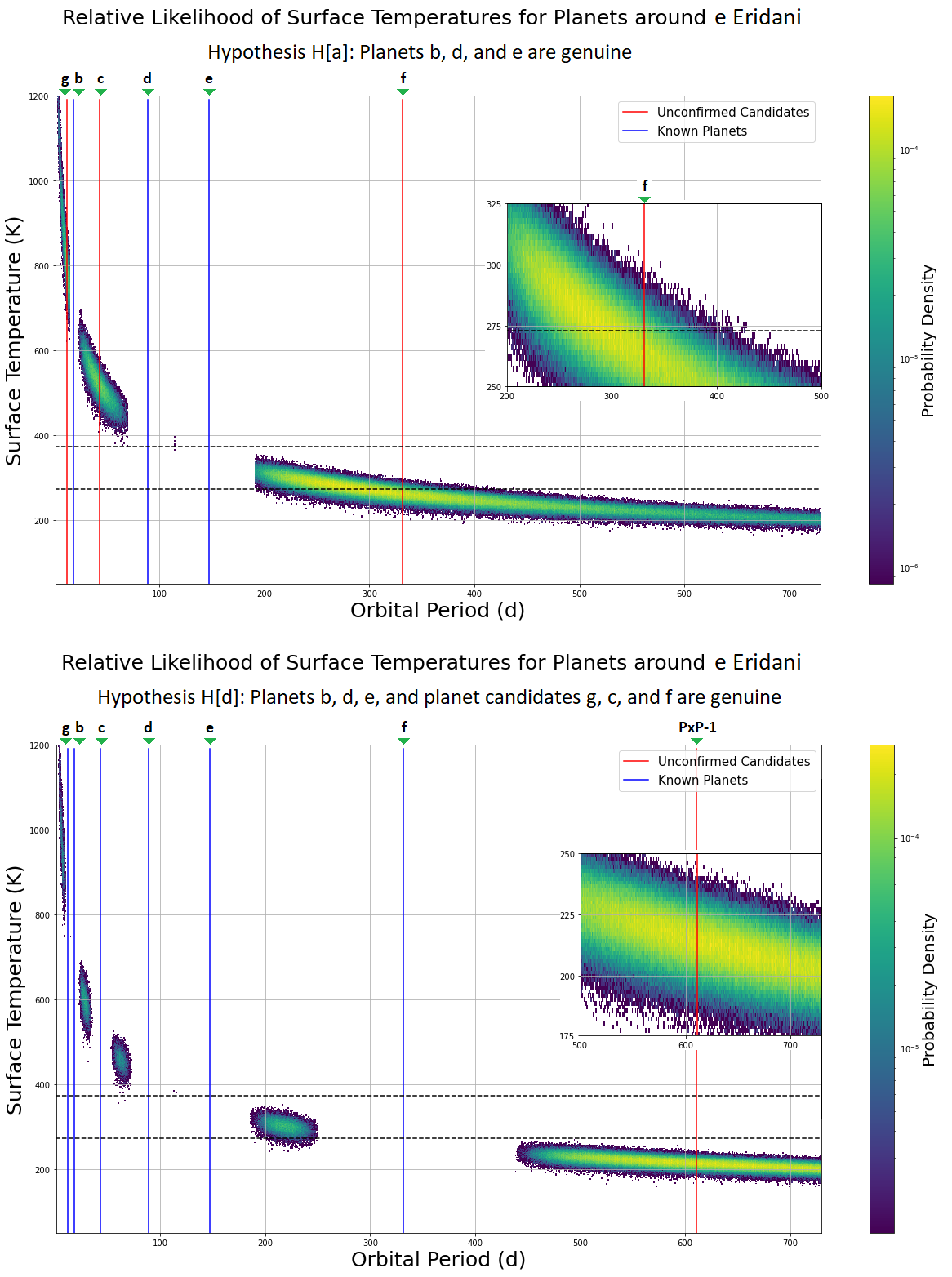}
    \caption{The upper plot shows the predictions of surface temperatures of \tnt{} predicted planets in the \eeri{} system for Hypothesis H[a].  The lower plot shows the predictions of surface temperature of \tnt{} predicted planet in the \eeri{} system for Hypothesis H[d]. Eccentricities for all planets in the system were set to zero for calculating these temperatures, but utilizing the expected low eccentricity values typical of these multi-planet systems does not significantly affect the temperature.}
    \label{fig:dynamitetemp}
\end{figure*}

The most likely surface temperature of planet e lies in the range of $330-358$ K, making it possible for planet e to harbor liquid water on its surface.  The surface temperature of planet candidate f, estimated to be in the range of $240-278$ K, is very similar to that of the Earth and makes it a candidate for harboring liquid water as well.  If planet f has a mass greater than $10 M_\oplus$, then based on the mass-radius relationship by \citet{Otegi2020}, the planet will likely be gaseous rather than rocky, making it uninhabitable.  We estimate that planet candidate g's surface temperature will be in the range of $788-882$ K, making its surface nearly as hot as that of Venus.  Planet candidate H[d]--PxP--1 would have a similar equilibrium and surface temperature to Mars, so if it is rocky and massive enough to maintain a significant atmosphere to increase the greenhouse effect, it could also be potentially habitable. The surface temperatures of planets and candidates are shown in Figure \ref{fig:dynamitetempcompare} and Figure \ref{fig:dynamitetemp}. In addition to applying our model to the \eeri{} planetary system -- as a simple test of the new temperature prediction module -- we also used it to estimate the surface temperature of Mars. Specifically, when Mars was hidden from the algorithm, based on the other inner Solar System planets, \tnt{} not only "predicted" Mars' orbit but also predicted a surface temperature to lie between $184-220$~K. While running the \tnt{} analysis, we excluded Jupiter from the inner Solar system and ran the simulation with Mercury, Venus, and Earth. In comparison, the measured temperature range is $165 - 235$~K \citep[][]{Martinez2017}. Thus, at least for planets without thick atmospheres and strong greenhouse effect, our approach provides a good initial estimate.
A more detailed description of what we did is given in Appendix~\ref{appendix: Mars} and the results are shown in Figure~\ref{fig:Mars}. 

Though our process for temperature estimates is simplistic, it is a reasonable approach given how little is known about the planets -- detected or predicted -- in the \eeri{} system.  As planet properties are refined in the future, the single-layer leaky greenhouse model used in our study should be replaced with more comprehensive models.

\subsection{Potential for Habitable Worlds} \label{subsec:potential_disc}

A key component of planetary surface habitability is the planet's ability to host stable liquid water on its surface.  Consequently, a simple constraint on the surface temperature of the planet might seem a valid first step towards characterizing a planet's habitability.  Surface temperatures of planet e and planet candidate f lie between 273 to 373 K, which makes them potentially suitable for hosting liquid water on their surfaces.  However, even after this constraint, the planet might be uninhabitable due to several factors like insufficient surface gravity for atmospheric retention, or accretion of massive $H_{2}-He$ envelope.

The minimum mass of planet candidate f, as determined by \citetalias[][]{Feng2017}, is $10.26^{+1.89}_{-1.47} M_{\oplus}$, which likely makes it a gaseous planet.  Thus, planet candidate f, even after lying close to the inner edge of the habitable zone of the \eeri{} system, might not be habitable.  \citet{2007A&A...476.1373S} demonstrated that a necessary, but insufficient on its own, condition for a planet to be potentially habitable is that the equilibrium temperature of the planet must be below $270$~K.  They further show that in such a case, if the surface temperature of the planet remains below the critical temperature of water ($T_{c} = 647$~K), the thermal emission of the planet will not exceed the greenhouse runaway limit of 300 $Wm^{-2}$. In case the equilibrium temperature exceeds 270~K, either the planet will have insignificant water or its surface temperature will exceed 1400~K.  In both cases, the planet will become uninhabitable, thus planet e (with an estimated equilibrium temperature of $320 \pm 12$) might not be potentially habitable.

As outlined by \citet{2011ApJ...736L..25K}, the equilibrium temperature of a planet near the outer edge of the habitable zone lies in the range 175 -- 200~K based on selective cloud coverage.  Thus, we believe that the newly predicted planet H[d]--PxP--1 at 611 days would be a potentially habitable planet candidate if it has the right mass, has an equilibrium temperature between $175 - 200$~K  and can even lie in the habitable zone as determined by \citet{2013ApJ...765..131K}. However, for this newly predicted planet to have the possibility of being habitable, it likely must have a significant amount of water, such that it can host liquid water for any temperature between 273 -- 647~K, and it likely must be geologically active so that $CO_{2}$ accumulates in the atmosphere as soon as the surface temperature drops below 273~K \citep{2007A&A...476.1373S}.

\section{Biosignature Studies with Future Telescopes and Missions} \label{sec:bio}

The combination of three factors make it very likely that the \eeri{} system will play pivotal roles in the future of exoplanet exploration.  First -- as our study highlights -- it hosts a rich planetary system with (possibly) six small planets.  Second, \eeri{} is among the closest Sun-like stars and the closest system with such a large number of discovered/predicted exoplanets.  Third, one -- perhaps two -- of its small planets would orbit in the habitable zone.  

The very proximity of \eeri{} makes it an ideal target for near-future direct imaging studies: its habitable zone extends from 0.15\arcsec{} to 0.23\arcsec{} in projection (see \S\ref{S:EEri}).  As a southern star (Declination $-$43$^\circ$), it is an ideal target for both the European Extremely Large Telescope's (E-ELT) PCS high-contrast, extreme adaptive optics imaging/spectrograph \citep[][]{KasperPCS2021}, as well as for the Giant Magellan Telescope's high-contrast imaging systems (e.g., GMagAO-X, \citealt{Males2019}).  With its habitable zone planets seen at about 160 mas separation from \eeri{}, these instruments will likely be able to detect it in visible/near-infrared reflected/scattered light.

The detectability of the planets within the habitable zone will depend, fundamentally, on whether the given telescope's inner working angle (IWA) will be smaller than the projected star--planet separation and, if so, whether the star--planet contrast will be detectable considering instrumental and astrophysical background signal. To explore which planets may be viable candidates for future instruments, in Table~\ref{tab:directimaging} we compare their maximum angular separation with the inner working angle typical to those instruments. For example, The LUVOIR mission concept's ECLIPS coronagraph would provide continuous coverage at wavelengths ranging from 200~nm to 2.0~\micron{} at a spectral resolving power of R=140 (in the visible) and R=70--200 (in the near-infrared). The inner working angle (IWA) of ECLIPS would be 4~$\lambda/D$ (in the ultraviolet) and 3.5 ~$\lambda$/D (in the visible and near-infrared). Coronagraphic imagers on HabEx \citep[][]{HabEx2020} and on large ground-based telescopes (e.g., \citealt{Males2019} and \citealt{KasperPCS2021}) will also allow very similar contrast-limited performance, but coupled to different diameter mirrors. Assuming a 4$\lambda/D$ limit, for each planet we calculate the required minimum mirror diameter at 0.4~\micron{} and at 1.5 $\micron{}$. These minimum diameters are shown in Table~\ref{tab:directimaging}. We determine the detectability (last three columns) by comparing the inner working angle for the given instrument to the maximum projected separation of the planet. Our assessment suggests that planets close to or within the habitable zone (f and H[d]-PxP-1) are excellent candidates for both space- and ground-based high-contrast imaging.

Furthermore, future thermal infrared imagers at the VLT -- perhaps building on the NEAR concept \citep[e.g.,][]{Kasper2017,Wagner2021} -- may also be  able to detect some of the planets close to the habitable zone of \eeri{}. In fact, \eeri{} has also been identified among the highest-priority targets for the high-resolution thermal infrared imaging with the E-ELT's METIS instrument \citep[][]{Bowens2021}.

\begin{table*}[ht]
    \centering
    \caption{Assessment of the planets and planet candidates for direct imaging potential.}
    \label{tab:directimaging}
    \begin{tabular}{|l|c|c|c|c|c|c|c|c|}
        \hline
        Planet & Period & Semi-major & Max. Proj. & \multicolumn{2}{c|}{Min. Mirror Diam. [m]}   &  \multicolumn{3}{c|}{Detectability}\\
        \cline{5-9}
        Name & [days] & axis [au] & Sep. [mas] & at 0.5~\micron{} & at 1.5~\micron{} &  ELTs &   LUVOIR & HabEx \\
        \hline
        g & $11.86_{-0.02}^{+0.01}$ & $0.095_{-0.001}^{+0.001}$ & 16 &  32 & 95 & N & N & N\\
        b & $18.33_{-0.02}^{+0.01}$ & $0.127_{-0.001}^{+0.001}$ & 21 & 24 & 72 & Y & N & N\\
        c & $43.17_{-0.10}^{+0.12}$ & $0.225_{-0.003}^{+0.002}$ & 37 & 14 & 40 & Y & Y & N\\
        d & $88.90_{-0.41}^{+0.37}$ & $0.364_{-0.004}^{+0.004}$ & 60 & 8.4 & 25 & Y & Y & N\\
        e & $147.02_{-0.91}^{+1.43}$ & $0.509_{-0.006}^{+0.006}$ & 84 & 6.0 & 18 & Y & Y & N\\
        f & $331.41_{-3.01}^{+5.08}$ & $0.875_{-0.010}^{+0.011}$ & 144 & 3.5 & 10 & Y & Y& Y\\
        H[d]-PxP-1 & 611 & 0.99. & 163 & 3.0 & 9 & Y & Y & Y \\
        \hline
    \end{tabular}
    \\[10pt]
    {Note}: The required minimum mirror diameter assumes the planet is observed at maximum elongation and assumes an inner working angle of 4.0 $\lambda$/D. 
\end{table*}

Finally, future interferometric missions focusing on habitable planets -- such as the LIFE mission concept \citep[][]{Quanz2021} -- will likely focus on \eeri{} due to its relatively easy-to-observe yet rich planetary system and likely presence of potentially habitable planets.

\section{Summary} \label{sec:summary}

We analyzed \eeri{} system for its multiplicity, the planetary architecture and orbital parameters of the planets, the equilibrium and surface temperature of the planets and potential candidates, and the dynamical stability of the system based on varying planetary eccentricities.

Starting from the assumption that the system consisted of only three planets -- b, d, and e -- the \tnt{} analysis predicted three more potential candidates -- g, c, and f -- which had also been reported in earlier studies but could not be confirmed. No information about the potential candidates was used as an input while running the \tnt{} analysis. The planetary parameters for these planet candidates were in good agreement with those reported in previous works (\citetalias[][]{Feng2017} and \citetalias[][]{Pepe2011}). Subsequently,  when all the 6 planets were used as an input, \tnt{} predicted one additional planet in the habitable zone of the system (P = 611 d) with a likely radius and mass of $1.25-5.16 R_\oplus$ and $1.03-8.7 M_\oplus$, respectively.

Our analysis of the equilibrium and surface temperatures of the planets was based on an albedo prior derived from the celestial bodies in the Solar system and the assumption of thin atmosphere. We find that though planet candidate f lies in habitable zone, its likely gaseous nature will make it uninhabitable. We find that the newly predicted planet (H[d]-PxP-1) might be habitable.

From our eccentricity analysis, we find that if \eeri{} is a three-planet system with planets b, d, and e, then the combined mean eccentricity for the system to be stable is $\sim0.05$. If the system is a six-planet system instead, then the combined mean eccentricity for the system to be stable is of order $\sim 0.026$. In either case, we find that the eccentricity of each planet should be significantly lower than the value fitted to the RV data, as also proposed by \citet[][]{Feng2017}.

As the planetary system's stability necessitates a lower-than-reported eccentricity for the planets, our analysis is based on this assumption. If better constraints on the eccentricities become available in future, then our analysis could be repeated again with the updated values. 

Our assessment of the \eeri{} system provides support for a seven-planet architecture with one planet candidate in the habitable zone of the system. Given its proximity and -- in projection -- large habitable zone as well as its rich inner planetary system, the nearby \eeri{} is likely to further ascend on the lists of planetary systems that are the most promising for exploration via direct imaging and for the search for life.

\acknowledgments
The results reported herein benefited from collaborations and/or information exchange within the program “Alien Earths” (supported by the National Aeronautics and Space Administration under Agreement No.  80NSSC21K0593) for NASA’s Nexus for Exoplanet System Science (NExSS) research coordination network sponsored by NASA’s Science Mission Directorate.  We acknowledge use of the software packages NumPy \citep{Harris2020}, SciPy \citep{Virtanen2020}, and Matplotlib \citep{Hunter2007}.  This paper includes data collected by the Kepler mission. Funding for the Kepler mission is provided by the NASA Science Mission Directorate.  We obtain the data set from the NASA Exoplanet Archive \citep{ps}\footnote{Accessed on 2022-01-23 at 09:00, returning 1 row.)} This dataset or service is made available by the NASA Exoplanet Science Institute at IPAC, which is operated by the California Institute of Technology under contract with the National Aeronautics and Space Administration. The citations in this paper have made use of NASA’s Astrophysics Data System Bibliographic Services.  This research uses data from the Solar System Dynamics Small-Body Database, which is hosted by the Jet Propulsion Laboratory and sponsored by NASA under Contract NAS7-030010.  This work has made use of data from the European Space Agency (ESA) mission {\it Gaia} (\url{https://www.cosmos.esa.int/gaia}), processed by the {\it Gaia} Data Processing and Analysis Consortium (DPAC,
\url{https://www.cosmos.esa.int/web/gaia/dpac/consortium}).  Funding for the DPAC has been provided by national institutions, in particular the institutions participating in the {\it Gaia} Multilateral Agreement.

\bibliography{main}

\appendix{}

\section{Selection of Bond Albedos for planets} \label{sec:albedoselection}

Our estimate of the equilibrium temperatures is informed and influenced by a prior Bond albedo distribution.  Though the albedo distribution of exoplanets is not yet determined, we can provide a reasonable prior on the Bond albedos considering current observational constraints on exoplanets and data on Solar System bodies.  \citet{2014ApJ...794..133S} devised a statistical method to estimate the average albedo of an ensemble of exoplanets.  They stacked multiple light curves of sub-Saturn planet candidates from the Kepler Object of Interest table, between $2R_\oplus - 6R_\oplus$, and measured the average depth of the light curve, from which they estimated the average albedo of the ensemble.  They considered two different cases: including the planet Kepler-10b and excluding it, such that Kepler-10b does not dominate the results due to its unusually deep secondary eclipse.

In a follow-up study, \citet{2017AJ....154..160S} expanded the sample size of the ensemble and divided the ensemble into three sub categories: $1-2R_\oplus$, $2- 4R_\oplus$, and $4-6R_\oplus$.  They then subdivided each class further into two different lists based on if the false-positive probability was less than $1\%$.  In addition, the analysis provided two cases, one case assuming that the planet completely redistributed the heat and the second assuming instantaneous re-radiation of heat by the planet.  \citet{2018MNRAS.478.3025J} used a different statistical approach for estimating the average albedo of Terran and Neptunian planet ensembles.  Instead of using light curves and planet candidates, multiple phase curves of confirmed planets were stacked, in order to estimate the albedos of the ensemble from the resulting average phase curve.  The resulting albedos from all these works can be found in Table \ref{tab:albedos}.

Solar System bodies offer greater amount of information and with higher precision than current exoplanet studies, albeit for objects that may differ in nature from the exoplanets such as those found in the \eeri{} system.  Nevertheless, we consider the approximate range and distribution of the Bond albedos for Solar System bodies as a general context for our prior.  Our discussion starts with asteroids and then continues with moons, rocky planets, and giant planets.  

\citet{2019A&A...626A..87S} calculated phase integral values of Solar System asteroids as a function of their geometric albedos and classified all the asteroids in three main classes: low albedo, medium albedo, and high albedo asteroids.  Bond albedo is defined as $A_{B} = A_g q$, where $A_g$ is the geometric albedo of the body and $q$ is the phase integral.  Therefore, the Bond albedos can be calculated for numerous asteroids listed in the dataset provided by the Jet Propulsion Laboratory through its Solar System Dynamics Small-Body Database\footnote{\url{https://ssd.jpl.nasa.gov/sbdb\_query.cgi}}.  On using the average phase integral values of the three asteroid classes, we find two discontinuities in the Bond Albedo distribution of asteroids.  To remove these discontinuities in the Bond Albedo distribution, instead of using the average phase integral for different classes, we use the linear fit equation, q = 0.359 + 0.47p \citep[][with the variables as above]{2019A&A...626A..87S} to calculate the phase integral, from which we calculate the Bond albedo of each asteroid.  Besides asteroids, we also collected Bond albedos of the eight planets and many satellites in our Solar System, which can be found in Table \ref{tab:planetalbedos}.

We compared the Bond albedos of various bodies in our Solar System with the constraints placed on Bond albedos by various studies (as presented in Table~\ref{tab:albedos}).  Venus might be an outlier as it is not consistent with the results from the studies by \citet{2014ApJ...794..133S}, \citet{2018MNRAS.478.3025J}, and \citet{2017AJ....154..160S}.  Earth's albedo is consistent with \citet{2018MNRAS.478.3025J}, but it is only consistent with the results of \citet{2017AJ....154..160S} under the assumptions of complete heat redistribution and include Kepler-10b, along with those of \citet{2014ApJ...794..133S} which exclude Kepler-10b.  Mars' albedo is almost consistent with \citet{2018MNRAS.478.3025J}.  Lastly, the albedos of Uranus and Neptune are consistent with both \citet{2018MNRAS.478.3025J} and \citet{2017AJ....154..160S}, but are only consistent with those results of \citet{2014ApJ...794..133S} which exclude Kepler-10b.  When using \tnt{} to predict the size of the planets around \eeri{}, we find that the radii of planets lie roughly in $1-2R_\oplus$ range, and likely to be Terran or Neptunian in nature.  As a result, we chose the upper limits placed on the Bond albedo by \citet{2018MNRAS.478.3025J}, as it is a broad range which meaningfully captures maximum data points and matches the likely natures of the \eeri{} planets.  With these upper limits, nearly 69\% and 99.8\% of albedos drawn from our adopted Gaussian distribution fall below $A_{B} = 0.35$ and $A_{B} = 0.63$.  Thus, we conclude that the Gaussian distribution of albedos is a reasonable prior for our initial, exploratory estimate of the equilibrium temperatures of the predicted planets in the system.

From these albedos, we calculated the equilibrium temperature of various bodies using two simple assumptions: the celestial bodies radiate as a blackbody, and they are in radiative equilibrium with their surroundings.  We also used the Leaky Greenhouse Model (see \S\ref{subsec: temp}) to calculate equilibrium temperature of those bodies which are larger than Mercury in size.  We show the Bond albedos and Equilibrium Temperatures of all these bodies in Figure \ref{fig:Heatmap}, as well as a histogram showing the Bond albedo distribution of various bodies.  For bodies of the size of Mercury or larger, the median Bond albedo is 0.3 and the standard deviation is 0.178.  On the right hand side of this figure, we show our adopted Gaussian Distribution of Bond Albedos centered at $\mu = 0.3$ with $\sigma = 0.1$.

\begin{table*}[ht]
    \centering
    \caption{Albedo Estimation for various ensemble of planets}
    \label{tab:albedos}
    \begin{tabular}{|l|c|c|c|}
        \hline
        \textbf{Reference} & \textbf{Planetary Radius [$R_\oplus$]} & \textbf{Bond Albedo} & \textbf{Notes} \\
        \hline
         \multirow{2}{*}{\cite{2014ApJ...794..133S}} & \multirow{2}{*}{$1-6$} & 0.$33 \pm 0.09$  & Excluding Kepler-10b\\
         & & $0.56 \pm 0.08$ & Including Kepler-10b \\
        \hline
        \multirow{16}{*}{\cite{2017AJ....154..160S}} & \multirow{8}{*}{$1-2$} & $0.166 \pm 0.09$ & Original list, Excluding Kepler-10b, $f = 1/4$ \\
        & & $<0.26$ & Original list, Excluding Kepler-10b, $f = 2/3$\\
        & & $0.166 \pm 0.09$ & Shortened list, Excluding Kepler-10b, $f = 1/4$\\
        & & $<0.26$ & Shortened list, Excluding Kepler-10b, $f = 2/3$ \\
        & & $0.29 \pm 0.06$ & Original list, Including Kepler-10b, $f = 1/4$\\
        & & $0.06 \pm 0.11$ & Original list, Including Kepler-10b, $f = 2/3$\\
        & & $0.29 \pm 0.06$ & Shortened list, Including Kepler-10b, $f = 1/4$\\
        & & $0.03^{+0.11}_{-0.12}$ & Shortened list, Including Kepler-10b, $f = 2/3$\\
        \cline{2--4}
        & \multirow{4}{*}{$2-4$} & $0.11 \pm 0.05$ & Original list, $f = 1/4$\\
        & & $< 0.06$ & Original list, $f = 2/3$\\
        & & $0.08 \pm 0.06$ & Shortened list, $f = 1/4$\\
        & & $<0.11$ & Shortened list, $f = 2/3$\\
        \cline{2--4}
        & \multirow{4}{*}{$4-6$} & $0.18 \pm 0.12$ & Original list, $f = 1/4$\\
        & & $0.14 \pm 0.12$ & Original list, $f = 2/3$\\
        & & $0.35 \pm 0.17$ & Shorted list, $f =1/4$\\
        & & $0.32 \pm 0.17$ & Shorted list, $f = 2/3$\\
        \hline
        \multirow{2}{*}{\cite{2018MNRAS.478.3025J}} & $0.61 - 1.25$ & $< 0.63$ & With $95 \%$ confidence \\
        & $1.28 - 5.47$ & $< 0.35$ & With $95 \%$ confidence \\
        \hline
    \end{tabular}
    \\[10pt]
\end{table*}

\begin{table*}[ht]
    \centering
    \caption{Albedos of planets, Satellites and Asteroids}
    \label{tab:planetalbedos}
    \begin{tabular}{|l|c|c|}
        \hline
        \textbf{Name} & \textbf{Bond Albedo} & \textbf{Reference}\\
        \hline
        Mercury & $0.088 \pm 0.003$ & \cite{2017arXiv170302670M} \\
        \hline
        Venus & 0.77 & \cite{2019AREPS..47..141J} \\
        \hline
        Earth & 0.3 & \cite{2021arXiv210302673H} \\
        \hline
        Mars & 0.25 & \cite{2017MNRAS.472....8Z} \\
        \hline
        Jupiter & $0.503 \pm 0.012$ & \cite{2018NatCo...9.3709L} \\
        \hline
        Saturn & 0.5 & \cite{2020ApJ...889...51M} \\
        \hline
        Uranus & $0.3 \pm 0.049$ & \multirow{2}{*}{\cite{2020RSPTA.37800474H}} \\
        \cline{1-2}
        Neptune & $0.29 \pm 0.067$ & \\
        \hline
        Moon & 0.12 & \cite{2017JGRE..122.2371H} \\
        \hline
        Pluto & $0.72 \pm 0.07$ & \cite{2017Icar..287..207B} \\
        \hline
        Charon & $0.29 \pm 0.05$ & \cite{2019ApJ...874L...3B} \\
        \hline
        Rhea & $0.48 \pm 0.09$ & \multirow{5}{*}{\cite{2010LPI....41.2035P}} \\
        \cline{1-2}
        Dione & $0.52 \pm 0.08$ & \\
        \cline{1-2}
        Tethys & $0.61 \pm 0.09$ & \\
        \cline{1-2}
        Mimas & $0.67 \pm 0.10$ & \\
        \cline{1-2}
        Enceladus & $0.85 \pm 0.11$ & \\
        \hline
        Io & 0.6 & \cite{2019ApJ...886..141M} \\
        \hline
        Portia Group & $0.026 \pm 0.005$ & \multirow{8}{*}{\cite{2001Icar..151...51K}}\\
        \cline{1-2}
        Puck & $0.035 \pm 0.006$ & \\
        \cline{1-2}
        Miranda & $0.2 \pm 0.03$ & \\
        \cline{1-2}
        Ariel & $0.23 \pm 0.025$ & \\
        \cline{1-2}
        Umbriel & $0.1 \pm 0.01$ & \\
        \cline{1-2}
        Titania & $0.17 \pm 0.015$ & \\
        \cline{1-2}
        Oberon & $0.14 \pm 0.015$ & \\
        \hline
        Phobos & $0.021 \pm 0.005$ & \cite{1998Icar..131...52S} \\
        \hline
        Deimos & $0.027 \pm 0.004$ & \cite{1996Icar..123..536T} \\
        \hline
        Callisto & 0.13 & \cite{1981Icar...46..137S} \\
        \hline
        Ganymede & 0.42 & \cite{2014Icar..243..429R} \\
        \hline
        Europa & $0.68 \pm 0.05$ & \cite{2016arXiv160807372A} \\
        \hline
        Titan & $0.27 \pm 0.02$ & \cite{1985Icar...62..425N} \\
        \hline
        Triton & $0.85 \pm 0.05$ & \cite{1991JGR....9619203H} \\
        \hline
        Phoebe & $0.023 \pm 0.007$ & \cite{2008Icar..193..309B} \\
        \hline
        Eris & 0.96 & \cite{2019MNRAS.489.2313S} \\
        \hline
        Comet 19P/Borelly & 0.018 & \cite{2007Icar..188..195L}\\
        \hline
        11 Parthenope & 0.07 & \cite{2011PASJ...63..499T} \\
        \hline
    \end{tabular}
    \\[10pt]
    \textbf{Note}: Albedos of Venus, Earth, Mars, Saturn, Moon, Io, Callisto, Ganymede, Eris, Borelly and 11 Parthenope did not have any uncertainity attached.
\end{table*}

\section{Additional Hypotheses} \label{sec: additionalhypotheses}

The four hypotheses presented in  \S\ref{sec:analysis} do not capture all the possible configurations of confirmed planets and unconfirmed planet candidates in the \eeri{} system. Therefore, we present here all the remaining possible combinations of the three confirmed planets and three planet candidates in the system, which is summarized in Table \ref{tab:additionalconfig}. As before, the starting point of our \tnt{} analysis for all scenarios are the three confirmed planets. When we use \tnt{} with planets b, d, and e, there are three strong likelihood predictions near the locations of the unconfirmed planet candidates g, c, and f. We then progress through each of the possible planet candidates, assuming one additional candidate is a genuine planet.  Finally, we show what the \tnt{} predictions give for the full 6-planet system.

\begin{table*}[ht]
    \centering
    \caption{Results for DYNAMITE analysis on additional configurations in the planetary system \eeri{}}
    \label{tab:additionalconfig}
    \begin{tabular}{|c|c|c|c|c|l|}
        \hline
        \textbf{Starting Point} & \textbf{First addition} & \textbf{Second Addition} & \textbf{Third Addition} & \textbf{Figure} & \textbf{Predictions} \\
        \hline
        \multirow{5}{*}{Planets b, d and e} & Planet candidate c & Planet candidate g & Planet candidate f & Figure~\ref{fig:bdecgf} & Strong prediction for a planet\\
        \cline{2-4}
        & \multirow{2}{*}{Planet candidate f} & Planet candidate c & Planet candidate g & Figure~\ref{fig:bdefcg} &  to exist at an orbital period \\
        \cline{3-4}
        & & Planet candidate g & Planet candidate c & Figure~\ref{fig:bdefgc} & of 611 days with a radius\\
        \cline{2-4}
        & \multirow{2}{*}{Planet candidate g} & Planet candidate c & Planet candidate f & Figure~\ref{fig:bdegcf} & of $1.1 - 2.0$ $R_\oplus$ and mass of\\
        \cline{3-4}
        & & Planet candidate f & Planet candidate c & Figure~\ref{fig:bdegfc} & $1.25 - 5.16$ $M_\oplus$.\\
        \hline
    \end{tabular}
    \\[10pt]
\end{table*}

\begin{figure*}[ht]
    \centering
    \includegraphics[width=0.7\columnwidth]{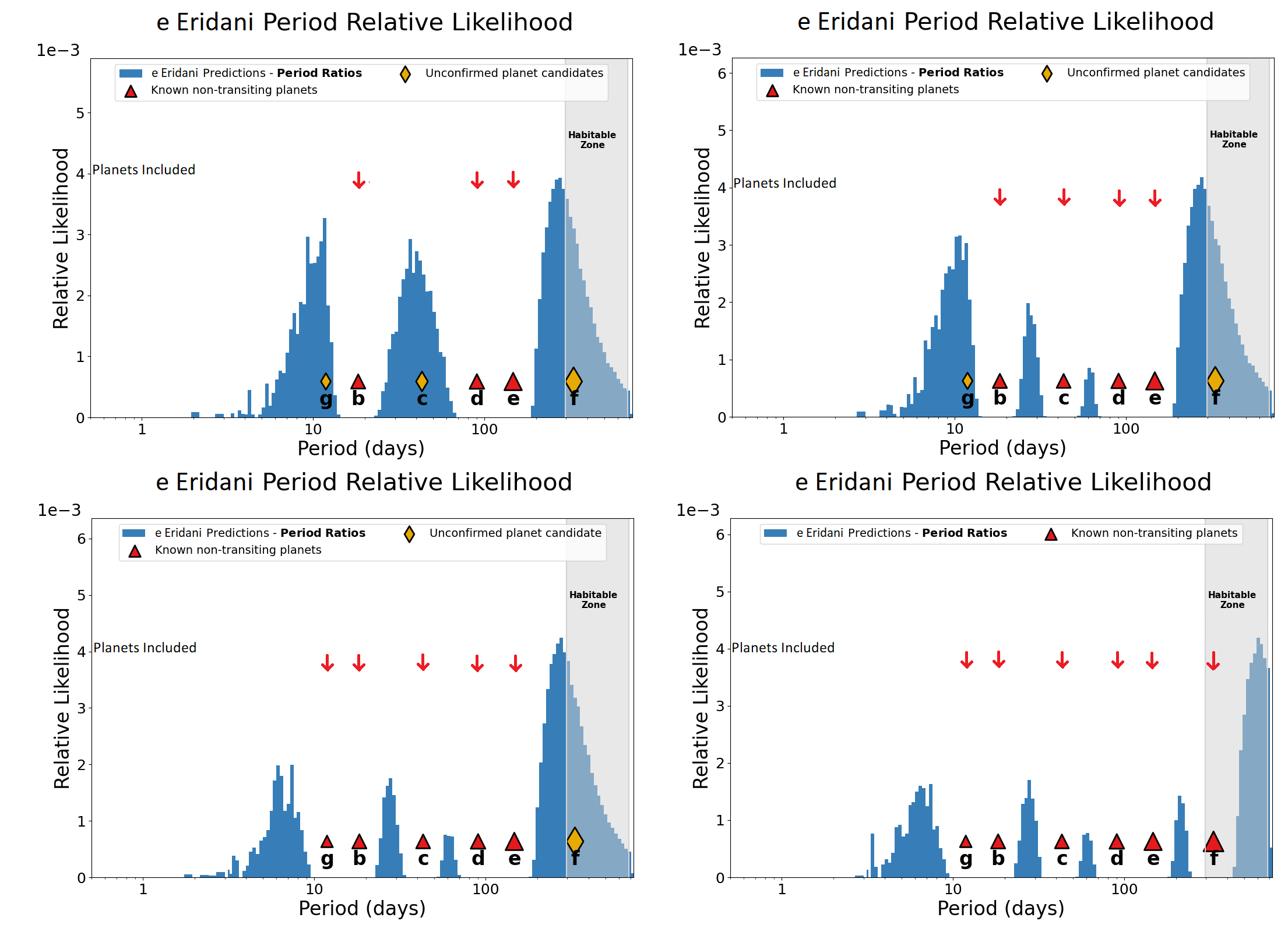}
    \caption{The upper left plot shows the \tnt{} analysis without planet candidates g, c, and f.  The upper right plot shows \tnt{} analysis without planet candidates g and f.  The lower left plot shows \tnt{} analysis without planet candidate f while the lower right plot shows the \tnt{} analysis with all the planets and planet candidates.}
    \label{fig:bdecgf}
\end{figure*}

\begin{figure*}[ht]
    \centering
    \includegraphics[width=0.7\columnwidth]{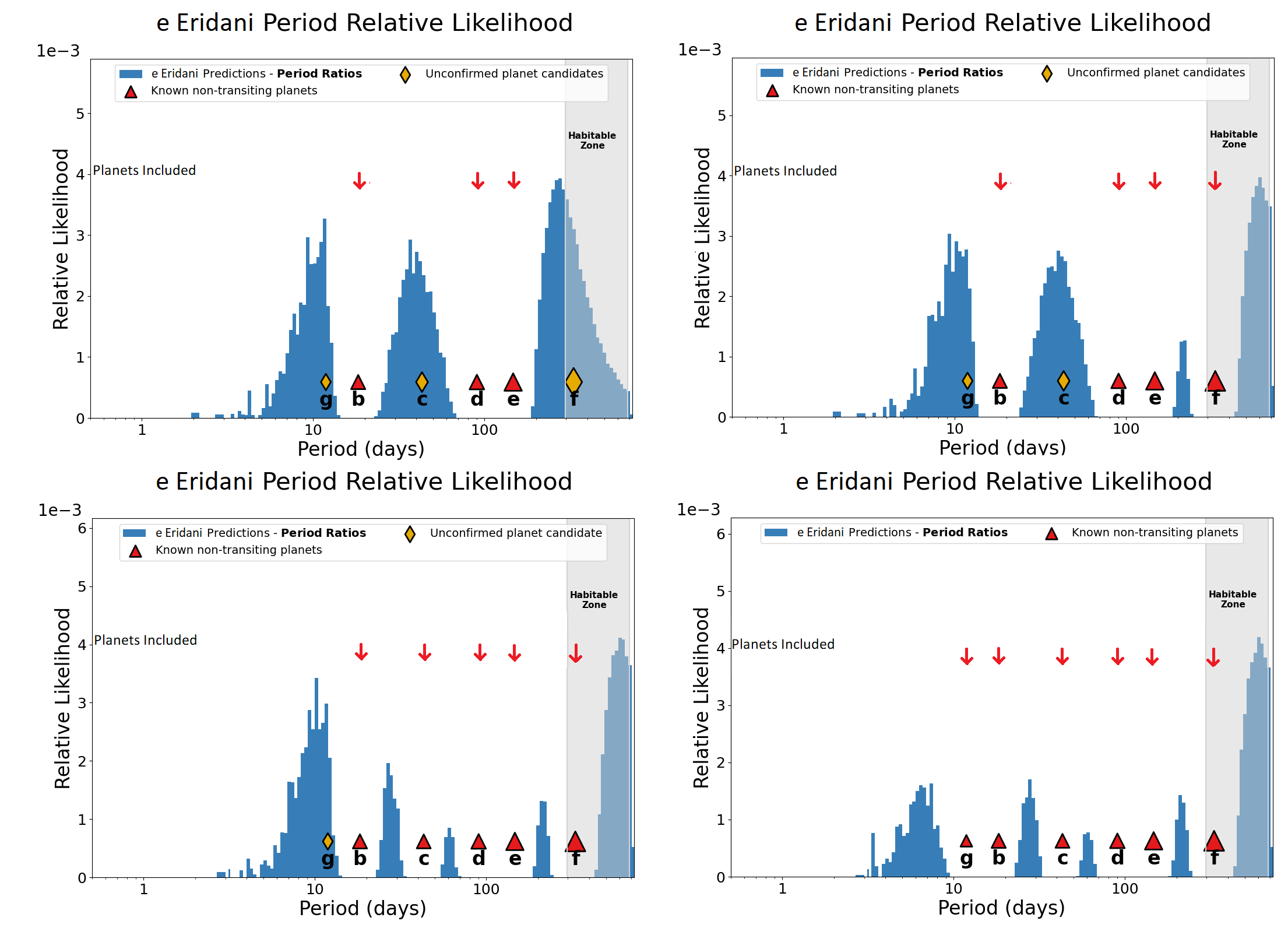}
    \caption{The upper left plot shows the \tnt{} analysis without planet candidates g, c, and f.  The upper right plot shows \tnt{} analysis without planet candidates g and c.  The lower left plot shows \tnt{} analysis without planet candidate g while the lower right plot shows the \tnt{} analysis with all the planets and planet candidates.}
    \label{fig:bdefcg}
\end{figure*}

\begin{figure*}[ht]
    \centering
    \includegraphics[width=0.7\columnwidth]{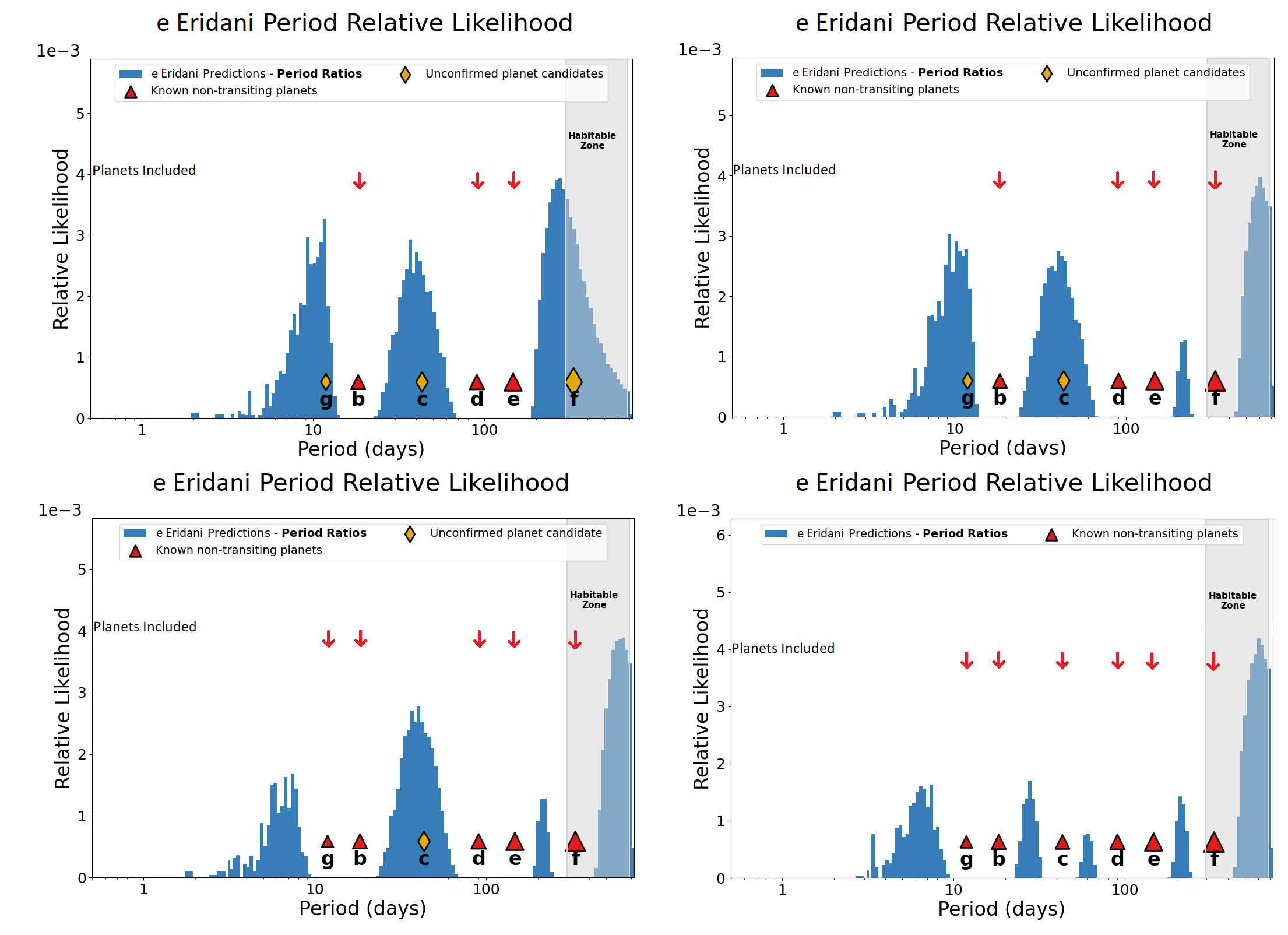}
    \caption{The upper left plot shows the \tnt{} analysis without planet candidates g, c, and f.  The upper right plot shows \tnt{} analysis without planet candidates g and c.  The lower left plot shows \tnt{} analysis without planet candidate c while the lower right plot shows the \tnt{} analysis with all the planets and planet candidates.}
    \label{fig:bdefgc}
\end{figure*}

\begin{figure*}[ht]
    \centering
    \includegraphics[width=0.7\columnwidth]{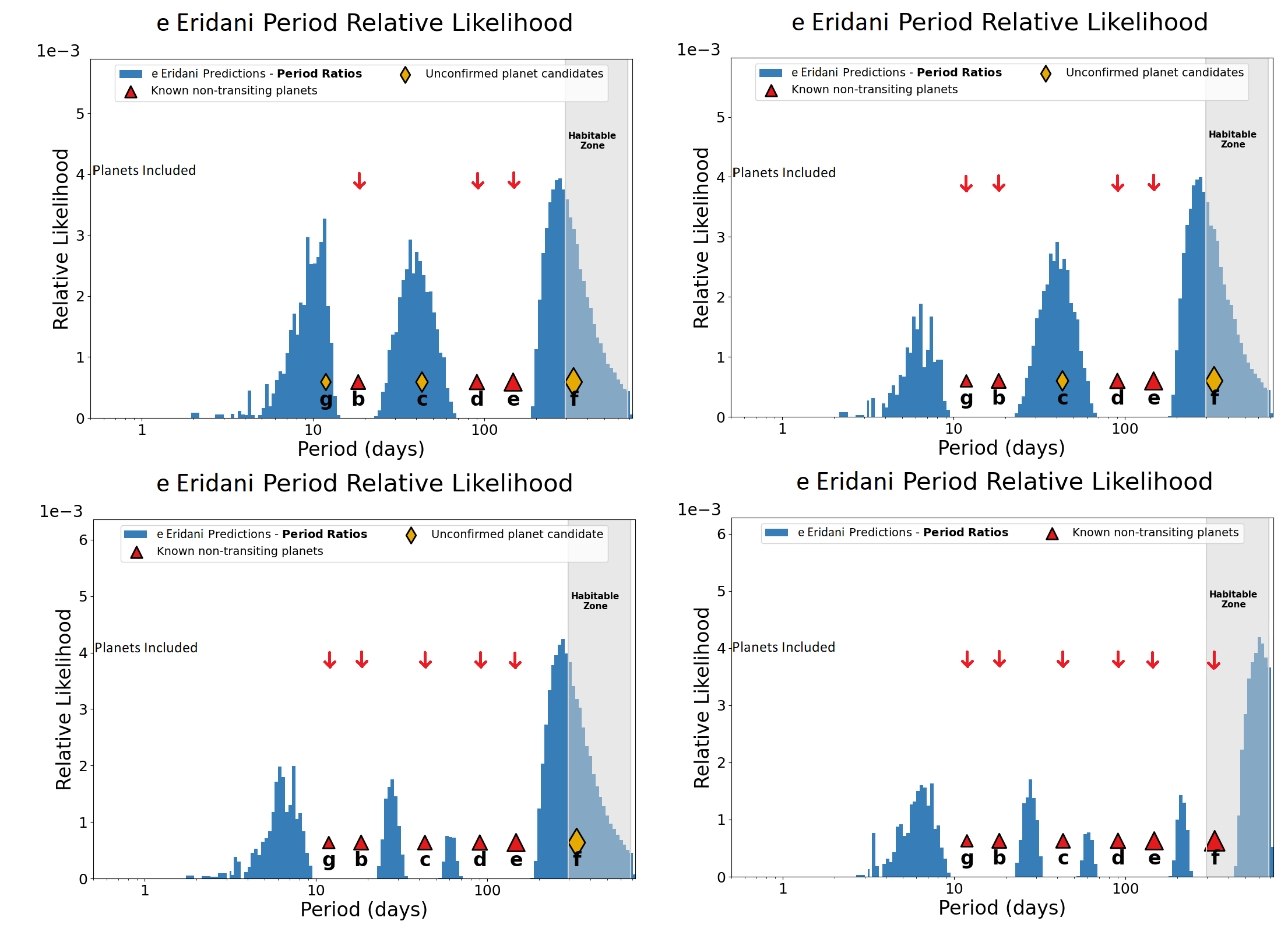}
    \caption{The upper left plot shows the \tnt{} analysis without planet candidates g, c, and f.  The upper right plot shows \tnt{} analysis without planet candidates c and f.  The lower left plot shows \tnt{} analysis without planet candidate f while the lower right plot shows the \tnt{} analysis with all the planets and planet candidates.}
    \label{fig:bdegcf}
\end{figure*}

\begin{figure*}[ht]
    \centering
    \includegraphics[width=0.7\columnwidth]{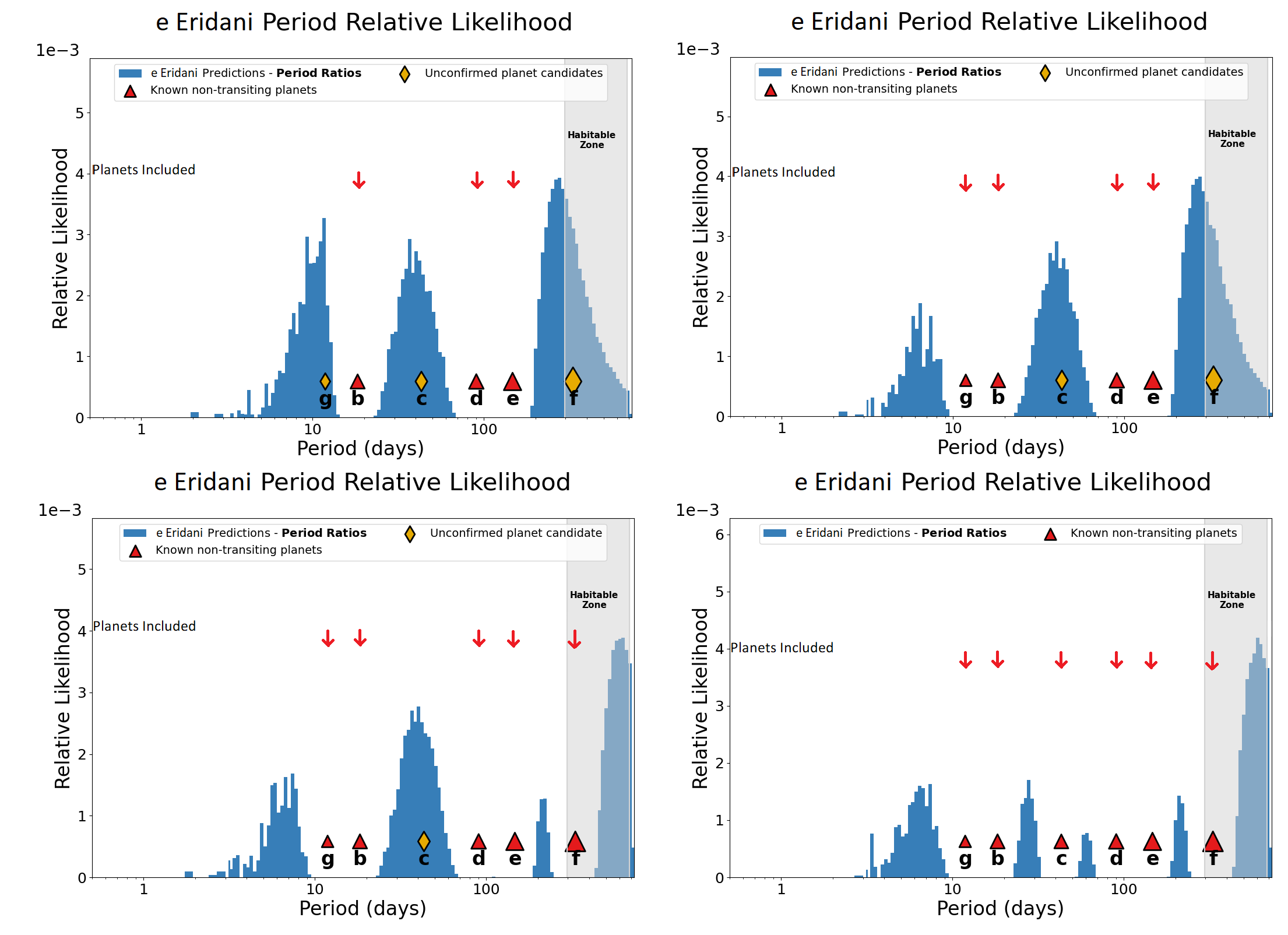}
    \caption{The upper left plot shows the \tnt{} analysis without planet candidates g, c, and f.  The upper right plot shows \tnt{} analysis without planet candidates c and f.  The lower left plot shows \tnt{} analysis without planet candidate c while the lower right plot shows the \tnt{} analysis with all the planets and planet candidates.}
    \label{fig:bdegfc}
\end{figure*}

\section{\tnt{} analysis on Inner Solar System and Temperature Estimation for Mars} \label{appendix: Mars}

In addition to analysing the \eeri{} planetary system, we also ran \tnt{} on the Inner Solar System while deliberately excluding Mars from the system. We found three peaks in the orbital period space distribution plot out of which the inner two peaks are significantly smaller than the outer peak. The first peak (at 43.5 days) lies interior to Mercury, the second peak (at 141 days) lies between Mercury and Earth, while the last peak (at 673 days) lies exterior to Earth. The outermost peak is less constrained as \tnt{} tries to match the Kepler period ratios by adding planets exterior to Earth. The outermost peak is the only significant peak and its location matches to that of Mars' orbital period ($\sim 687$ days -- an excellent 2\% match). This gives another line of evidence that the model accurately retrieves the intentionally removed planet. 

Using the methods explained earlier in \S\ref{sec:tempmethods}, we calculated the likelihood surface temperature of Mars to be $202 \pm 18$~K ($\pm$2$\sigma$ range approx. 166~K -- 238~K). Keeping in mind that Mars has a fairly complex climate and shows weak Greenhouse effect, our calculated surface temperature for Mars is an excellent match with the measured temperature range of the Martian surface $165-235$~K \citep{Martinez2017}. Therefore, the predicted surface temperature is in excellent agreement with the actual surface temperature range. Given the typical limitations on the atmospheric characterization of exoplanets, our model can provide useful initial exploratory estimates of the surface temperature ranges that may be typical to them. 

\begin{figure*}[ht]
    \centering
    \includegraphics[width=0.95\columnwidth]{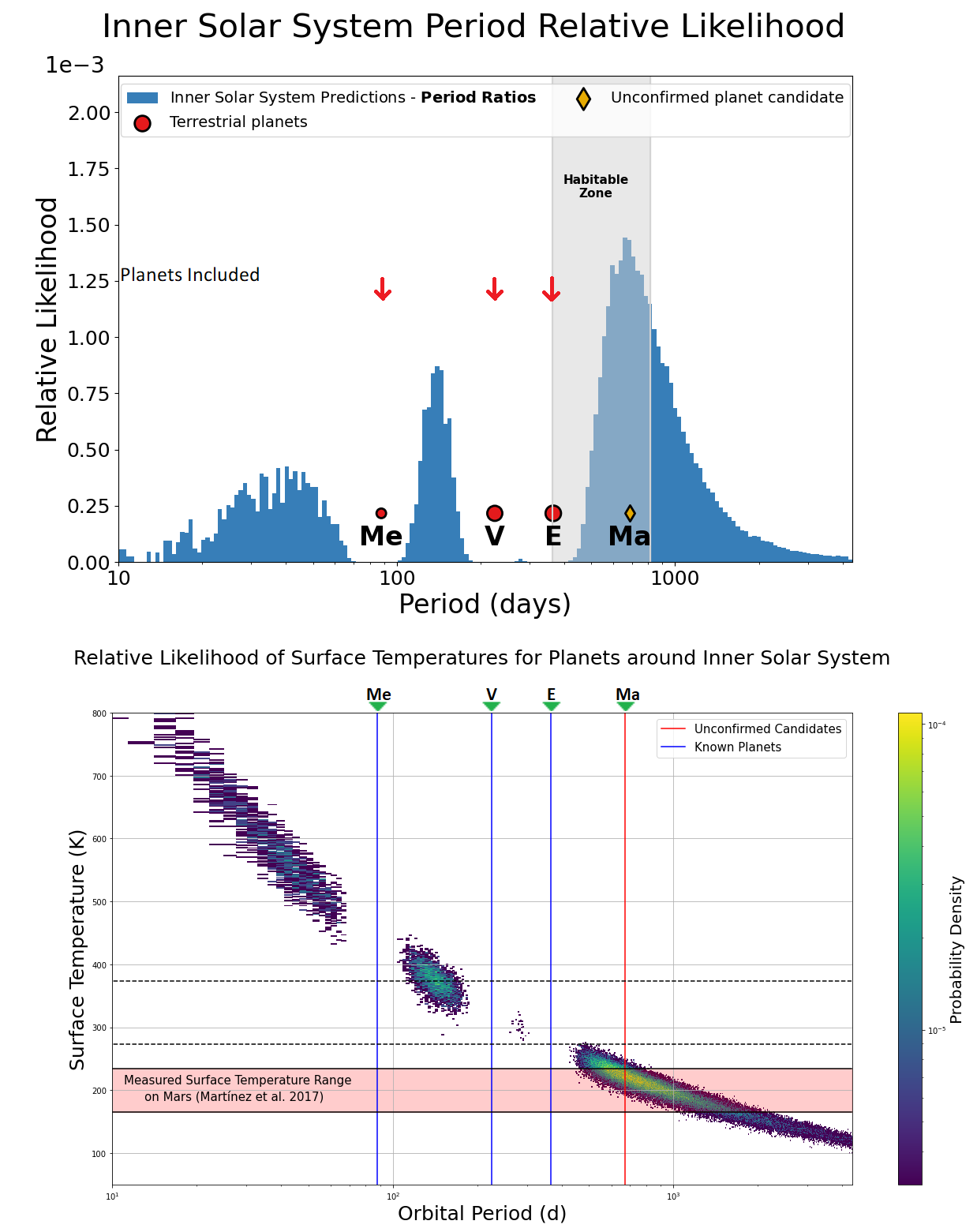}
    \caption{The upper plot shows \tnt{} analysis on Inner Solar system when Mars was excluded from the system. The lower plot shows surface temperature distribution for Inner Solar System based on \tnt{} analysis.}
    \label{fig:Mars}
\end{figure*}

\section{Dynamical Stability analysis of the e Eridani System}\label{appendix:eccana2}

\begin{figure*}[ht]
    \centering
    \includegraphics[width=0.9\linewidth]{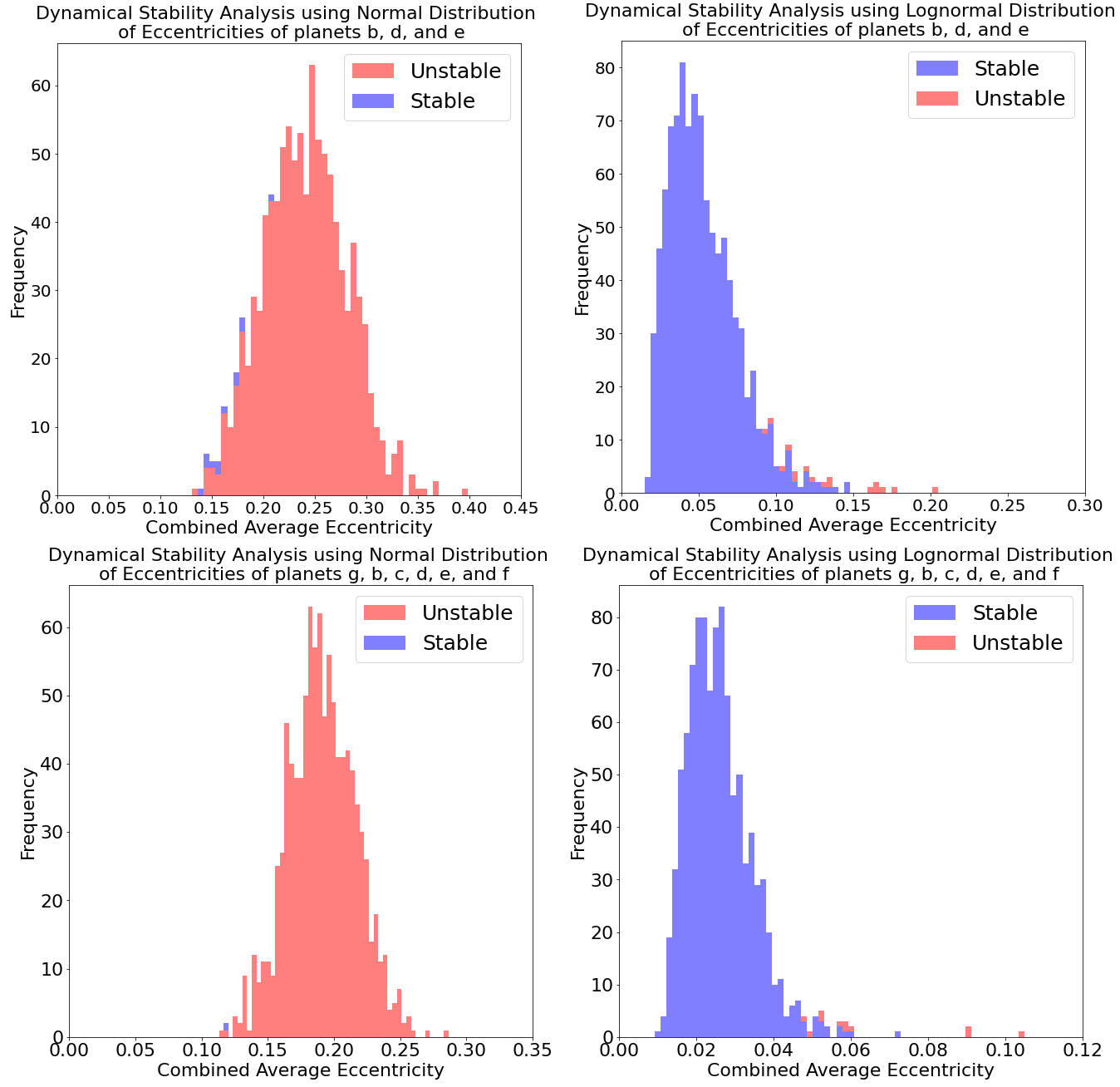}
    \caption{The number of stable vs unstable iterations in the four dynamic stability simulations - for the three- and six-planet systems assuming Normal and Lognormal eccentricity distributions.}
    \label{fig:ecc_ana}
\end{figure*}

Here we present the results of our dynamical stability analysis of \eeri{} planetary system as a function planetary eccentricities. We analyzed the three-planet and six-planet system with: (1) Normal eccentricity distribution following the RV solutions from \citet{Feng2017} and (2) Lognormal eccentricity distributions from \citet{2020AJ....160..276H} (see Section~\ref{subsec:eccentricityanalysis} for more details). We calculated the combined average eccentricity for each of the four cases. The results for each run are shown in Figure~\ref{fig:ecc_ana}.

\end{document}